\documentclass[preprint,showpacs,preprintnumbers,amssymb,aps]{revtex4}
\usepackage{graphicx}
\usepackage{dcolumn}
\usepackage{amsmath}
\usepackage{epsfig}
\usepackage{subfigure}
\usepackage{array}
\usepackage{epstopdf} 
\usepackage{xcolor}

\begin{document}

\normalsize

\title{\boldmath An Updated Study for $\Upsilon$ Production and Polarization at the Tevatron and LHC}

\author{
\small Yu Feng, Bin Gong, Lu-Ping Wan and Jian-Xiong Wang \\
 {\it Institute of High Energy Physics, Chinese Academy of Sciences, P.O.Box 918(4), Beijing 100049, China.
 }}

\begin{abstract}
  Following the nonrelativistic QCD factorization scheme, by taking latest available measurement on $\chi_b(3P)$
  into consideration, we present an updated study on
  the yield and polarization of $\Upsilon(1S,2S,3S)$ hadroproduction, and the fractions of $\chi_b(mP)$
  feed-down in $\Upsilon(nS)$ production at QCD next-to-leading order. In the fitting, three schemes are applied with
  different choice of $\chi_b(mP)$ feed-down ratios and NRQCD factorization scale.
  The results can explain the measurements
  on yield very well as in our previous work. The polarization puzzle to $\Upsilon(3S)$ is now solved by
  considering the $\chi_b(3P)$ feed-down contributions.
  The ratio of $\sigma[\chi_{b2}(1P)]/\sigma[\chi_{b1}(1P)]$ measurements from CMS can also be reproduced in our prediction.
  Among the different schemes, the results show little difference, but there are sizeable difference for the fitted
  long-distance color-octet matrix elements. It may bring large uncertainty when the values are applied in other
  experiments such as in $ee,~ep$ colliders.
\end{abstract}

\pacs{12.38.Bx, 13.60.Le, 13.88.+e, 14.40.Pq}

\maketitle

\section{Introduction}
Quantum chromodynamics(QCD) successfully describes the strong interaction at parton level due to its asymptotic freedom property.
But it fails to calculate observations with detected hadrons directly since the hadronization from quark is nonperturbative.
Therefore, factorization scheme to bridge the perturbative calculable part and nonperturbative hadronization part is crucial.
For heavy quarkonium production and decay, non-relativistic QCD(NRQCD) factorization scheme\cite{Braaten:1994vv,Bodwin:1994jh},
which was proposed to explain the huge discrepancy between the theoretical prediction and experimental measurement on the transverse
momentum distribution of $J/\psi$ production at the Tevatron, has been a very successful scheme in many applications.
However, it encounter
challenges on the transverse momentum distribution of polarization for $J/\psi$ and $\Upsilon$ hadroproduction
where the theoretical predictions can not describe the experimental measurements at QCD leading order (LO), or next-to-leading order (NLO).

In the last few years, significant progresses have been made in NLO QCD calculations.
The NLO corrections to color-singlet $J/\psi$ hadroproduction was investigated in Refs.~\cite{Campbell:2007ws,Gong:2008sn}, where
the $p_t$ distribution is found to be enhanced by 2-3 orders of magnitude at the high $p_t$ region and the $J/\psi$ polarization
changes from transverse into longitudinal at NLO~\cite{Gong:2008sn}. And the results are reproduced at LO in a new
factorization scheme for large $p_t$ quarkonium production~\cite{Kang:2011mg}.
In Ref.~\cite{Gong:2008ft}, NLO corrections to $J/\psi$ via S-wave color-octet(CO)
states($^1S_0^{[8]}$,~$^3S_1^{[8]}$)are studied and the $p_t$ distributions of both $J/\psi$ yield and polarization
changed little compared with LO.
The NLO corrections for $\chi_{cJ}$ hadroproduction are studied in Ref.~\cite{Ma:2010vd}.
Calculations and fits for both yield and polarization of $J/\psi$ production at NLO QCD are presented by three
groups~\cite{Butenschoen:2012px,Chao:2012iv,Gong:2012ug}.
Where the complete prompt $J/\psi$ hadroproduction study, which is corresponding to the present available experimental measurements,
is performed in Ref.~\cite{Gong:2012ug} for the first time.
The $J/\psi$ polarization puzzle is still not fully understood. Recently, the LHCb Collaboration~\cite{Aaij:2014bga} published their measurement of $\eta_c$
production. Three letters~\cite{Butenschoen:2014dra, Han:2014jya, Zhang:2014ybe} came out successively, investigating the data from different points of views.
Ref.~\cite{Butenschoen:2014dra} considered the $\eta_c$ experiment as a challenge of NRQCD, while Refs.~\cite{Han:2014jya, Zhang:2014ybe} emphasized its
indications on the $J/\psi$ productions and polarizations. The complicated situation suggests that, further study and phenomenological test of NRQCD is still
an urgent task.

$\Upsilon$ production and polarization is an alternative one of the best laboratories for understanding the physics in the hadronization of
heavy quark pairs. Due to its heavier mass and smaller $v$ (where $v$ is the velocity of the heavy quark
in the meson rest frame), one can expect better convergence in the QCD and NRQCD expansions, and
consequently better description of the experiment by QCD NLO predictions.
For $\Upsilon$ hadroproduction, similar progresses are also achieved on the $p_t$ distribution of yield
and polarization for the CS channel at QCD NLO~\cite{Campbell:2007ws,Gong:2008sn}, and for S-wave CO states~\cite{Gong:2010bk}.
The NLO correction via all CO state is studied in Ref.~\cite{Wang:2012is}.
The first complete NLO QCD corrections on yield and polarization of $\Upsilon(1S,2S,3S)$ are presented in our former work~\cite{Gong:2013qka},
where the results can explain the experiments on yield very well, so as the polarization of $\Upsilon(1S,2S)$ at
CMS measurements. However, without considering the $\chi_b(3P)$ feed-down contribution,
the polarization of $\Upsilon(3S)$ is inconsistent with data.
Thereafter, the mass of $\chi_b(3P)$ is measured~\cite{Aaij:2014hla} at the LHC,
and the fractions for $\Upsilon(3S)$ production from $\chi_b(3P)$ radiative decay is first
measured~\cite{Aaij:2014caa} by the LHCb Collaboration.
The large measured value of this fraction indicates that the reconsideration for $\Upsilon(3S)$,
as well as $\Upsilon(1S,2S)$, is needed.
In this work, we take into consideration of $\chi_b(3P)$ feed-down contributions carefully to update the yield and polarization analysis on
$\Upsilon(nS)$ hadroproduction at QCD NLO correction, and
also predict the ratio of differential cross sections of $\chi_{b2}(1P)$ to that of $\chi_{b1}(1P)$
on the LHCb~\cite{Aaij:2014hla} and CMS data~\cite{Khachatryan:2014ofa}.

This paper is organized as follows. In Sec.~\ref{charp:2}, we give a brief description of the theory in our work.
Numerical results are presented in Sec.~\ref{charp:3}. In Sec.~\ref{charp:4}, we introduce more fitting schemes and discuss the difference between them.
Finally, the summary and discussions are given in the last Section.

\section{Theory description}\label{charp:2}

Following the NRQCD factorization formalism~\cite{Bodwin:1994jh}, the cross section for quarkonium hadroproduction $H$
can be expressed as
\begin{eqnarray}
  d\sigma[pp\rightarrow H + X]=\sum_{i,j,n}\int dx_1 dx_2 G^i_p G^j_p
  \times d\hat{\sigma}[ij\rightarrow (b\overline{b})_n X]\langle{\cal O}^{H}_n \rangle
\end{eqnarray}
where $p$ is either a proton or an antiproton, $G^{i(j)}_p$ is the parton distribution function(PDF) of $p$,
the indices $i,j$ runs over all possible partonic spices,
and $n$ denotes the $b\overline{b}$ intermediate states ($^3S_1^{[1]}$,$^1S_0^{[8]}$,$^3S_1^{[8]}$,$^3P_J^{[8]}$) for
$\Upsilon$ and ($^3P_J^{[1]}$,$^3S_1^{[8]}$) for $\chi_{bJ}$. The short-distance contribution $d\hat{\sigma}$ can
be calculated perturbatively, while the long-distance matrix elements(LDMEs) $\langle{\cal O}^{H}_n \rangle$
are governed by nonperturbative QCD effects.

The polarizations of $\Upsilon$ are defined as~\cite{Beneke:1998re}
\begin{equation}
\lambda=\frac{d\sigma_{11}-d\sigma_{00}}{d\sigma_{11}+d\sigma_{00}},~
\mu = \frac{\sqrt{2}Re d\sigma_{10}}{d\sigma_{11}+d\sigma_{00}},~
\nu = \frac{\sqrt{2}Re d\sigma_{1,-1}}{d\sigma_{11}+d\sigma_{00}}.
\end{equation}
where $d\sigma_{S_zS'_z}$ is the spin density matrix of $\Upsilon$ hadroproduction.
Only the parameter $\lambda$ in helicity frame is considered in our work.

The fractions of $\Upsilon(nS)$ originating from $\chi_b(mP)$ decays is defined as
\begin{equation}
  {\cal R}^{\chi_b(mp)}_{\Upsilon(nS)}\equiv \sum_{J=0,1,2}
  \frac{\sigma(pp\rightarrow\chi_{bJ}(mP)X)}{\sigma(pp\rightarrow\Upsilon(nS)X)}\times {\cal B}[\chi_b(mP)\rightarrow \Upsilon(nS)]
\end{equation}
where $n$ and $m$ are radial quantum numbers of the bound states and ${\cal B}$ denotes the branching
ratios for the decay $\chi_{bJ}(mP)\rightarrow\Upsilon(nS)\gamma$.

To obtain $d\sigma_{S_zS'_z}$, similar treatment as in Ref.~\cite{Gong:2012ug,Gong:2013qka} is taken
for both direct and feed-down contributions.
For various feed-down contributions in $\Upsilon$ production, we treat them in different ways to get comparable results.
The details are given in the next section.

The newly updated Feynman Diagram Calculation package~\cite{Wang:2004du,Wan:2014vka} and the  are used in our calculation.
Compared with our former work, the newly added calculations are about the productions on $\chi_b(3P)$
in different experimental conditions.

\section{Numerical results}\label{charp:3}
In the numerical calculation, the CTEQ6M parton distribution functions~\cite{Pumplin:2002vw} and corresponding
two-loop QCD coupling constants $\alpha_s$ are used.
We adopt an approximation $m_b=M_H/2$ for the quark mass, where $M_H$ is the mass of bottomonium $H$.
All the masses are taken from PDG ~\cite{Beringer:1900zz}, except for $\chi_{bJ}(3P)$,
which are chosen as $M_{\chi_{bJ}(3P)}=$10.511 GeV for $J$=0,1,2~\cite{Aaij:2014caa}.
Therefore, the mass of bottom quark in our calculation is different values for $\Upsilon(nS)$ and $\chi_b(nP)$.

The color-singlet LDMEs are estimated from wave functions at the origin
\begin{eqnarray}
\nonumber\langle{\cal O}^{\Upsilon(nS)}(^{3}S^{[1]}_{1})\rangle&=&\frac{9}{2\pi}|R_{\Upsilon(nS)}(0)|^{2}, \\
 \langle{\cal O}^{\chi_{bJ}(mP)}(^{3}P^{[1]}_{J})\rangle&=&\frac{3}{4\pi}(2J+1)|R'_{\chi_{bJ}(mP)}(0)|^{2}.
\end{eqnarray}
while the wave functions and their derives via potential model calculation~\cite{Eichten:1995ch}.
We listed the results in Table~\ref{tab:potential}.

\begin{table}
\begin{center}
\caption{ \label{tab:potential}  Radial wave functions at the origin~\cite{Eichten:1995ch}.}
\footnotesize
\begin{tabular*}{80mm}{c@{\extracolsep{\fill}}ccc}
  \hline\hline
  $\Upsilon(nS)$ & $|R_{\Upsilon(nS)}(0)|^{2}$ & $\chi_b(mP)$ & $|R'{\chi_{b(mP)}}(0)|^{2}$ \\[0.1cm]
  \hline
  1S & 6.477$GeV^{3}$ & 1P & 1.417$GeV^{5}$ \\
  2S & 3.234$GeV^{3}$ & 2P & 1.653$GeV^{5}$ \\
  3S & 2.474$GeV^{3}$ & 3P & 1.794$GeV^{5}$ \\
  \hline\hline
\end{tabular*}
\end{center}
\end{table}

Branching ratios involving bottomonia can be found in Ref.~\cite{Gong:2013qka} Table I,
in which however, that for $\chi_{bJ}(3P)$ are not included.
Since no experimental data for branching ratios of $\chi_b(3P)$ feed-down to $\Upsilon(nS)$
is available right now, we take
${\cal B}[\chi_{bJ}(3P)\rightarrow\Upsilon(3S)] \simeq {\cal B}[\chi_{bJ}(2P)\rightarrow\Upsilon(2S)]$
as an approximation and ignored the contributions from $\chi_{bJ}(3P)$ for $\Upsilon(2S)$ and $\Upsilon(1S)$
due to the little fractions.

The factorization, renormalization and NRQCD scales are chosen as $\mu_f$ = $\mu_r$ = $\sqrt{4m^2_b+p^2_t}$
and $\mu_{\Lambda}$ = $m_b v \approx$ 1.5 GeV, respectively.
The center-of-mass energy is 1.8 TeV and 1.96 TeV for Tevatron Run I and Run II and, 7 TeV and 8 TeV for the LHC,
respectively. A shift $p^H_t\approx p^{H'}_t \times (M_H/M_{H'})$ is used while considering the kinematics effect
in the feed-down from higher excited states.

In the fit, we have used three kinds of data from experimental measurements, namely,
the differential cross section from CDF~\cite{Acosta:2001gv}, LHCb~\cite{LHCb:2012aa},
CMS~\cite{Khachatryan:2010zg} and ATLAS~\cite{Aad:2012dlq};
the polarization from CDF~\cite{CDF:2011ag} and CMS~\cite{Chatrchyan:2012woa};
the fractions of $\Upsilon(nS)$ production originating from
radiative decays of $\chi_b(mP)$ meson (${\cal R}^{\chi_b(mp)}_{\Upsilon(nS)}$)~\cite{Aaij:2014caa}.
We included the data of the fraction from $\chi_b(3P)$ feed-down at both
$\sqrt{s}$=7 TeV and $\sqrt{s}$=8 TeV in the fit of $\Upsilon(3S)$, while for $\Upsilon(2S)$ and $\Upsilon(1S)$,
only the points at $\sqrt{s}$=7 TeV for fractions are included.
The linear interpolation method has been taken when dealing with the fraction data
since it behaves smoothly and the $p_t$ points nearly our theoretical calculations are chosen.
Only the data in the region $p_t>8$ GeV are used in our fit as we know that the double expansion in $\alpha_s$ and $v^2$ is not good in the small $p_t$ regions.

We perform the fits for $\Upsilon(3S)$,$\Upsilon(2S)$,$\Upsilon(1S)$ hadroproduction step by step,
and the corresponding $\chi^2/d.o.f$ are 97/72,~114/47,~73/44.
All the fitted CO LDMEs are presented in Table~\ref{tab:LDMEs-def}.
A covariant-matrix method~\cite{Gong:2012ug} is performed for the plots in order to express the
uncertainty from the CO LDMEs properly.
But we only rotate the direct three LDMEs, namely,
$\langle{\cal O}^{\Upsilon(nS)}(^{1}S^{8}_{0})\rangle$,
$\langle{\cal O}^{\Upsilon(nS)}(^{3}S^{8}_{1})\rangle$,$\langle{\cal O}^{\Upsilon(nS)}(^{3}P^{8}_{J})\rangle$,
while the last one $\langle{\cal O}^{\chi_b(nP)}(^{3}S^{8}_{1})\rangle$ is fixed.
The descriptions for yield, polarization and fractions are given in the following subsections.

\begin{table}
\begin{center}
\caption{ \label{tab:LDMEs-def}The obtained CO LDMEs for bottomonia production(in units of 10$^{-2}$ GeV$^3$).
  $\chi_b(3P)$ feed-down contributions are only considered for $\Upsilon(3S)$ with branching
  ratios ${\cal B}[\chi_{bJ}(3P)\rightarrow\Upsilon(3S)] \simeq {\cal B}[\chi_{bJ}(2P)\rightarrow\Upsilon(2S)]$
  as an approximation. }
\footnotesize
\begin{tabular*}{160mm}{@{\extracolsep{\fill}}cccccc}
\hline\hline
  state ~&~ $\langle{\cal O}^{\Upsilon(nS)}(^{1}S^{[8]}_{0})\rangle$ ~&~
  $\langle{\cal O}^{\Upsilon(nS)}(^{3}S^{[8]}_{1})\rangle$ ~&~
  $\langle{\cal O}^{\Upsilon(nS)}(^{3}P^{[8]}_{0})\rangle/m_b^2$ ~&~
  state ~&~ $\langle{\cal O}^{\chi_b(mP)}(^{3}S^{[8]}_{1})\rangle$ \\ [0.1cm]
  \hline
  $\Upsilon(1S)$ ~&~ 13.6 $\pm$ 2.43 ~&~ 0.61 $\pm$ 0.24 ~&~ -0.93 $\pm$ 0.5  ~&~ $\chi_b(1P)$ ~&~ 0.94 $\pm$ 0.06\\
  $\Upsilon(2S)$ ~&~ 0.62 $\pm$ 1.98 ~&~ 2.22 $\pm$ 0.24 ~&~ -0.13 $\pm$ 0.43 ~&~ $\chi_b(2P)$ ~&~ 1.09 $\pm$ 0.14\\
  $\Upsilon(3S)$ ~&~ 1.45 $\pm$ 1.16 ~&~ 1.32 $\pm$ 0.20 ~&~ -0.27 $\pm$ 0.25 ~&~ $\chi_b(3P)$ ~&~ 0.69 $\pm$ 0.14\\
\hline\hline
\end{tabular*}
\end{center}
\end{table}

\subsection{Yield and polarization}

\begin{figure} [!ht]
  \centering
  \includegraphics[width=5.0cm]{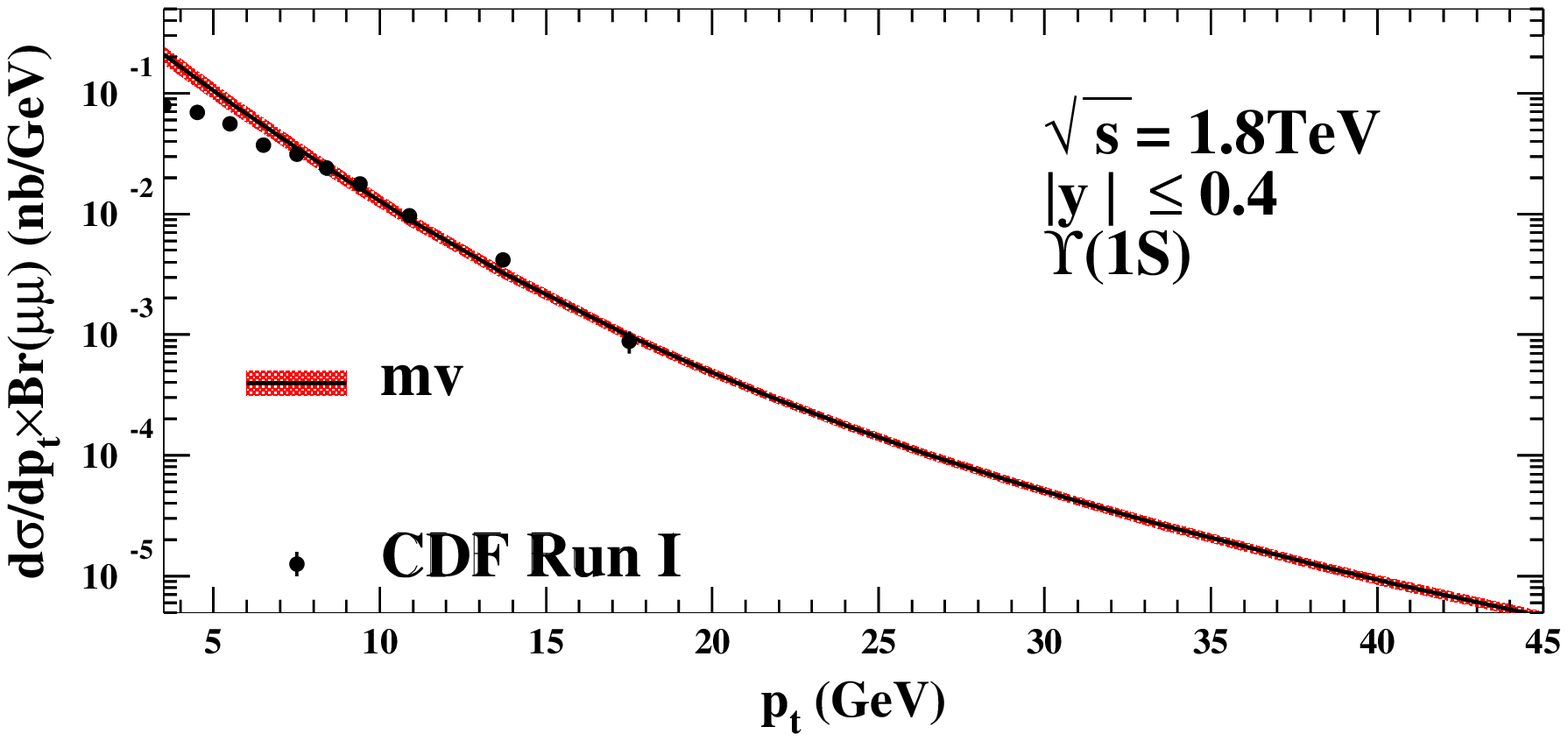}  \includegraphics[width=5.0cm]{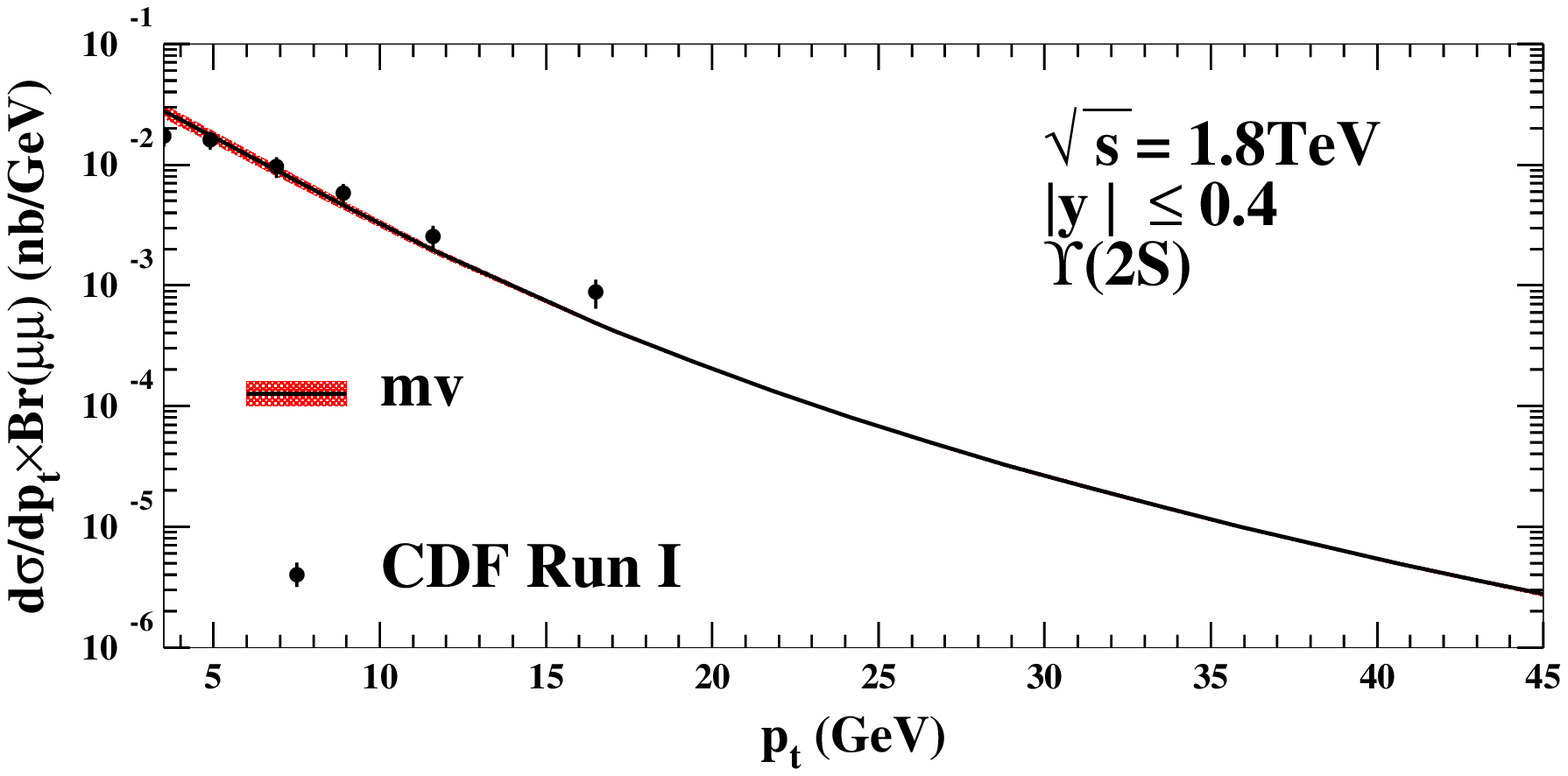} \includegraphics[width=5.0cm]{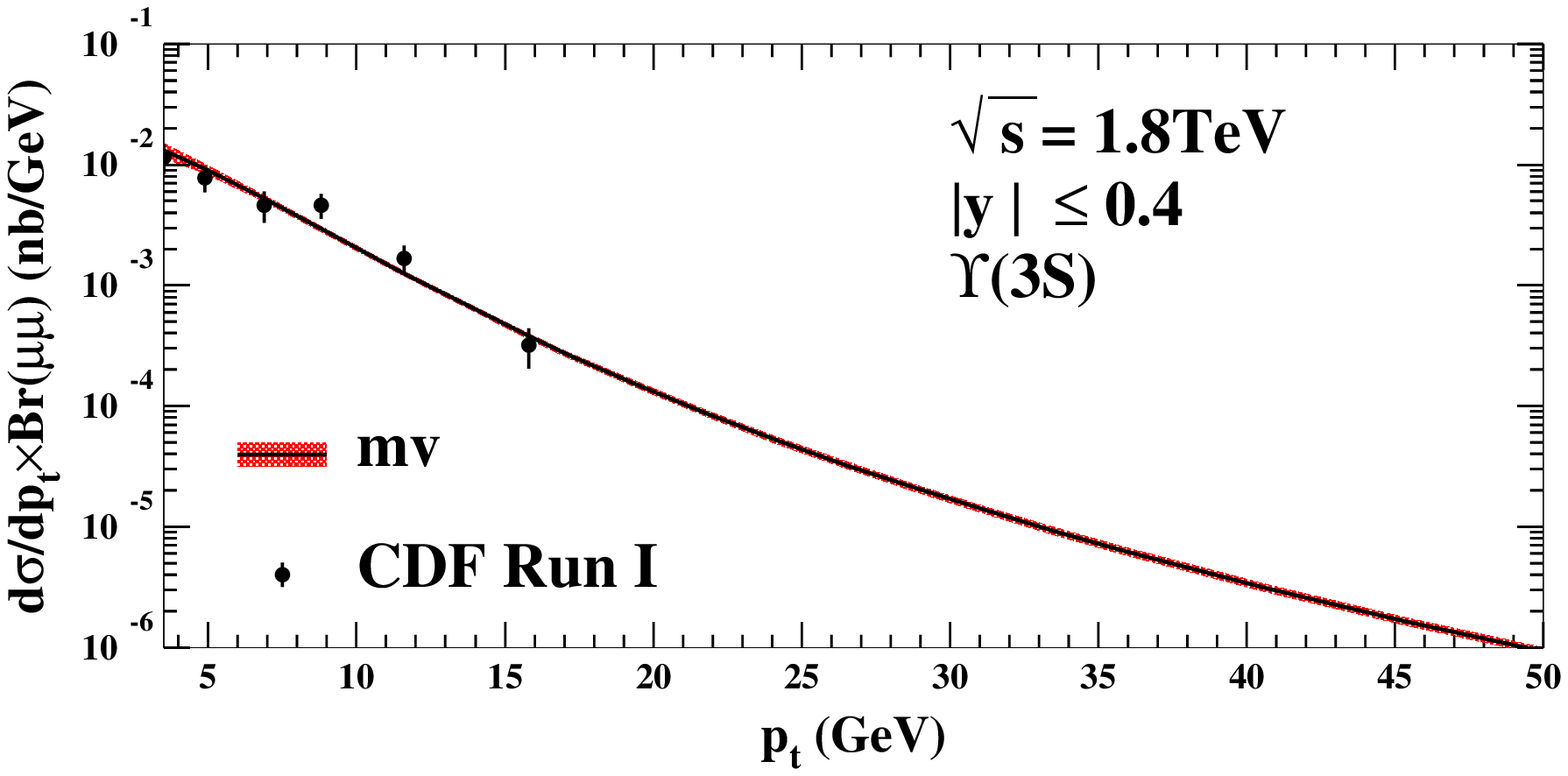} \\
  \includegraphics[width=5.0cm]{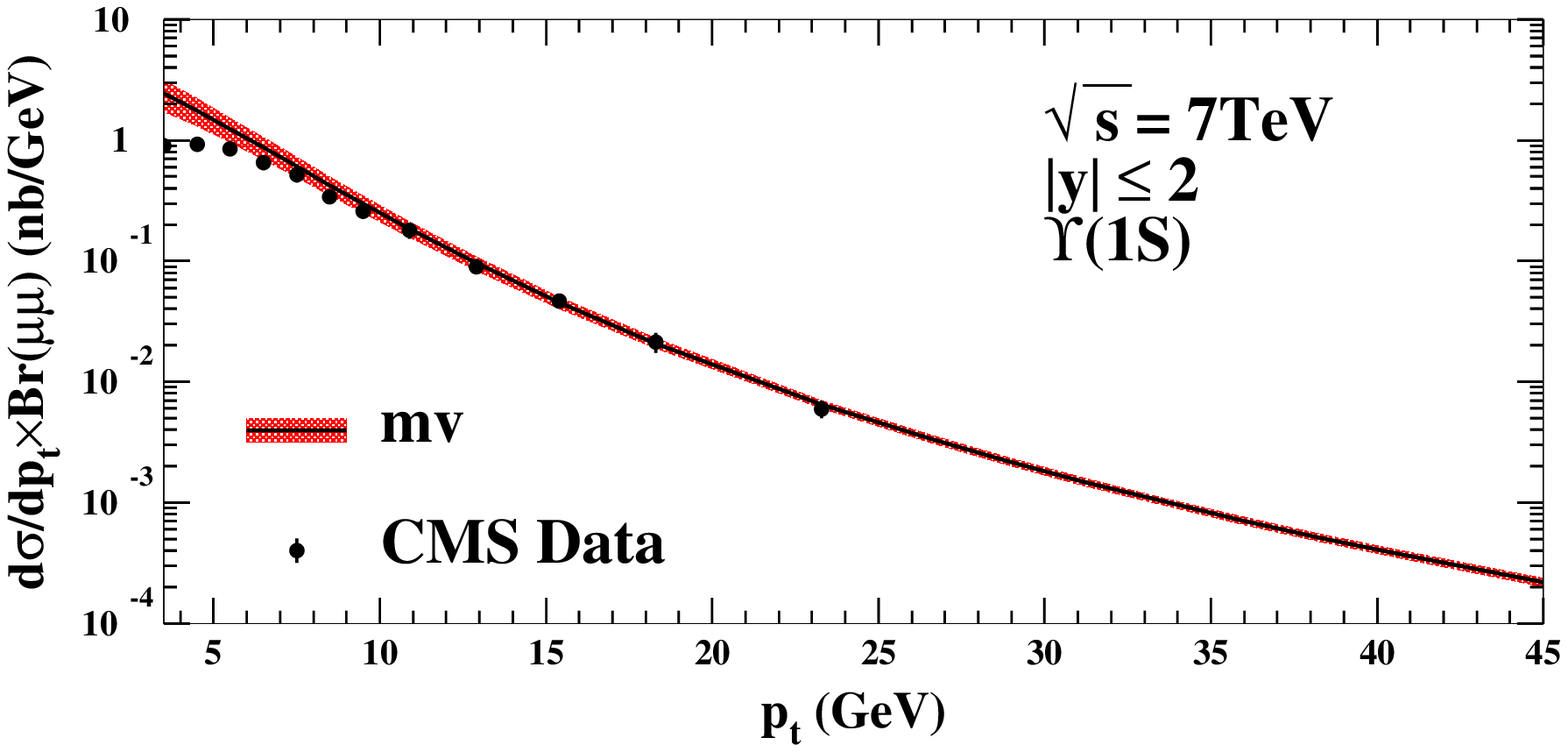}  \includegraphics[width=5.0cm]{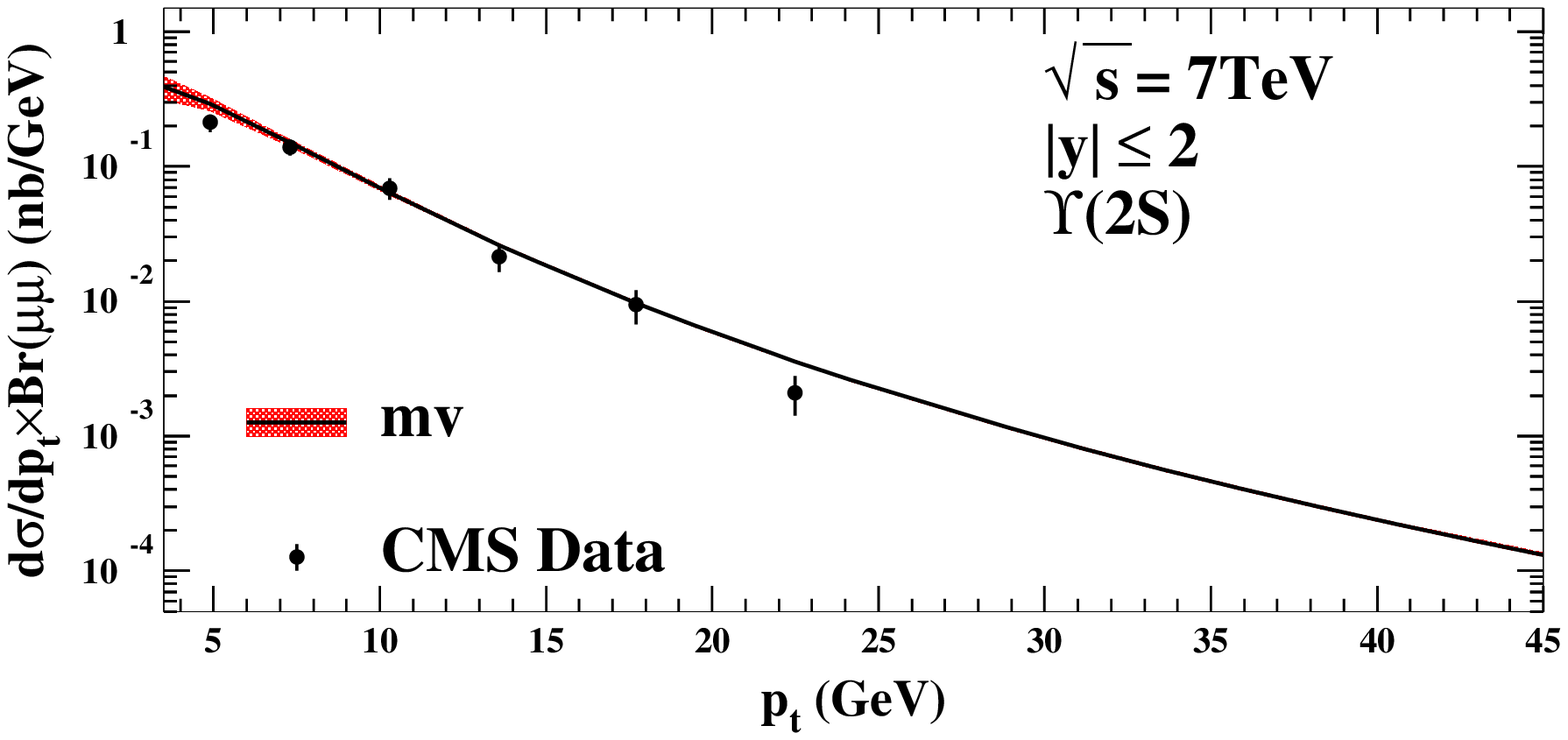} \includegraphics[width=5.0cm]{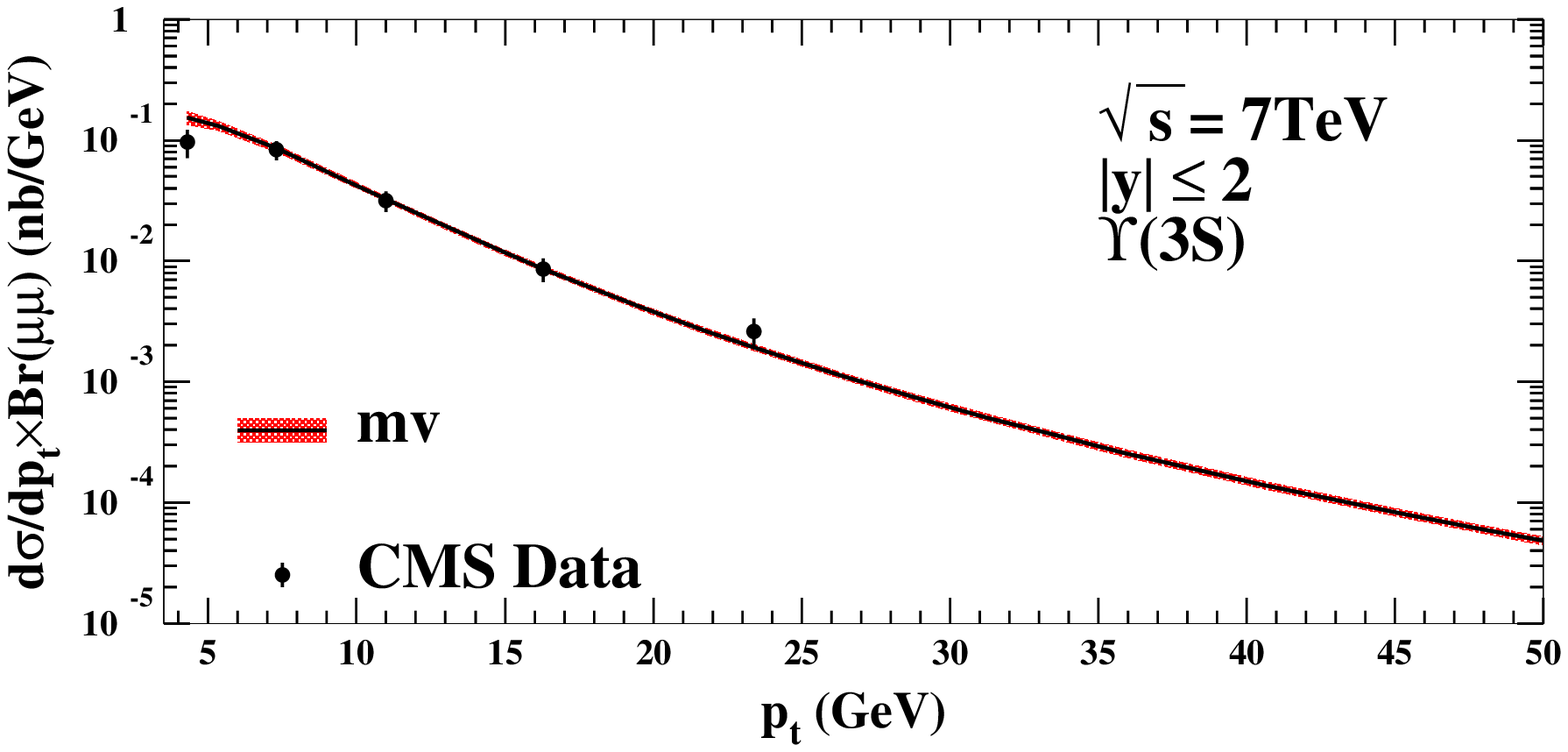} \\
  \includegraphics[width=5.0cm]{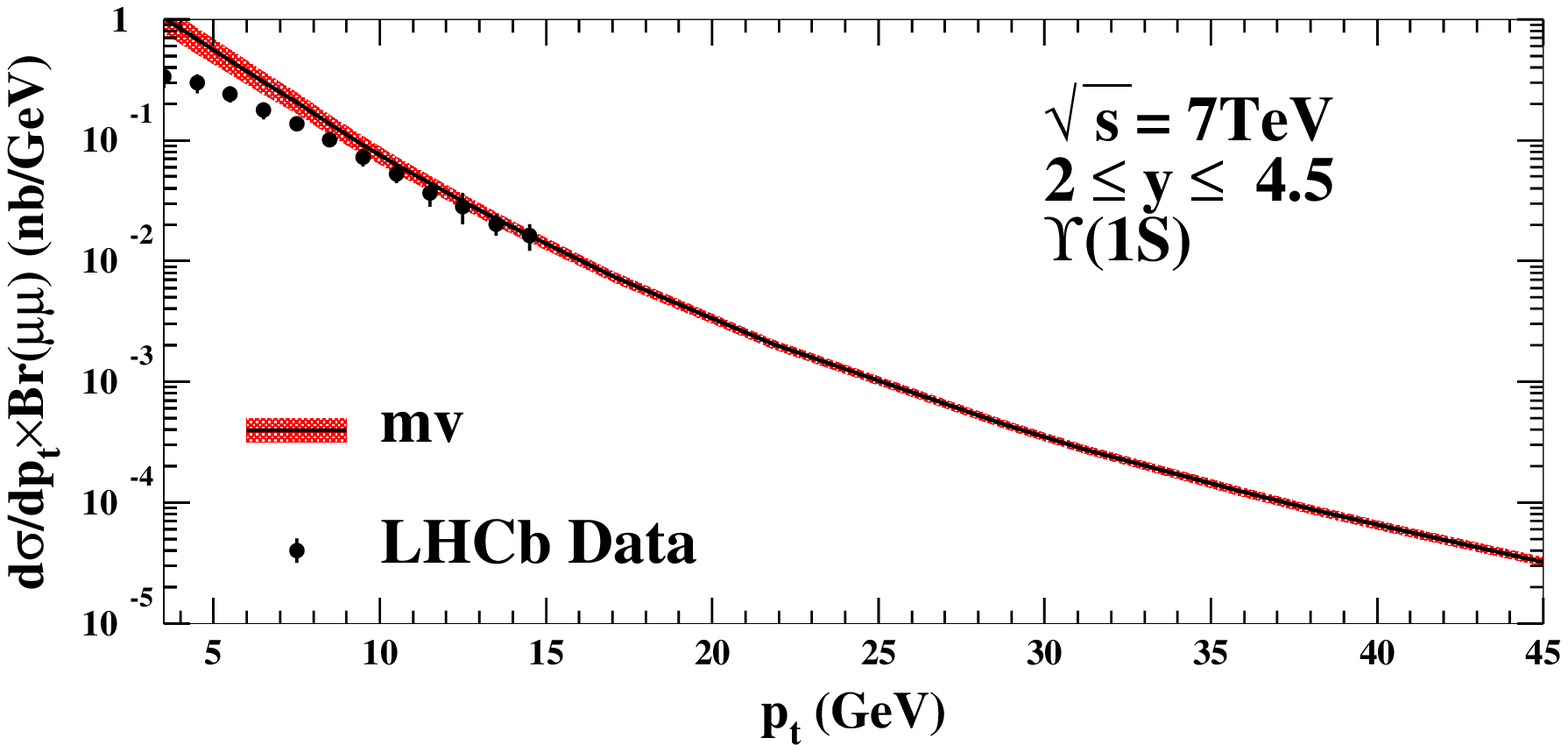}  \includegraphics[width=5.0cm]{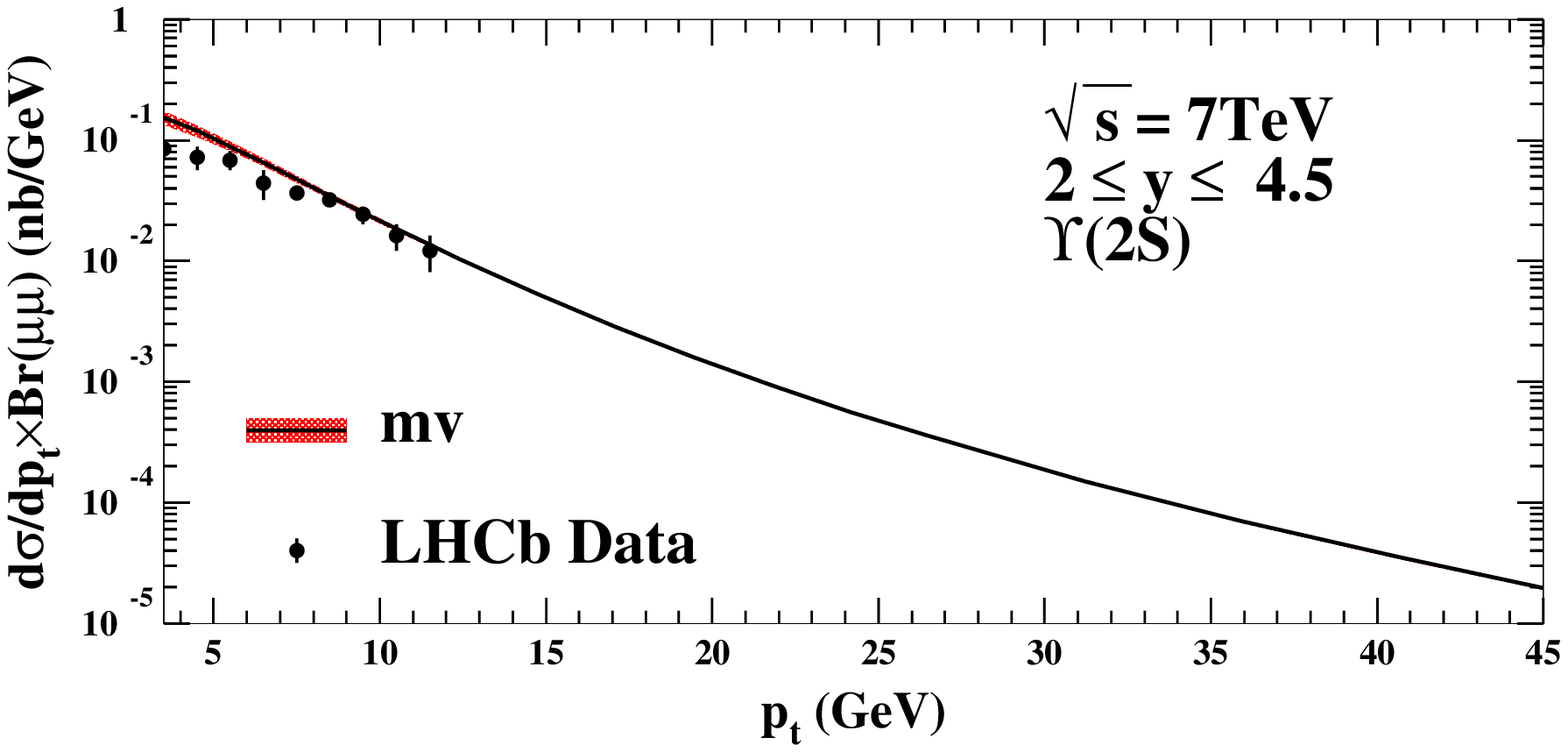} \includegraphics[width=5.0cm]{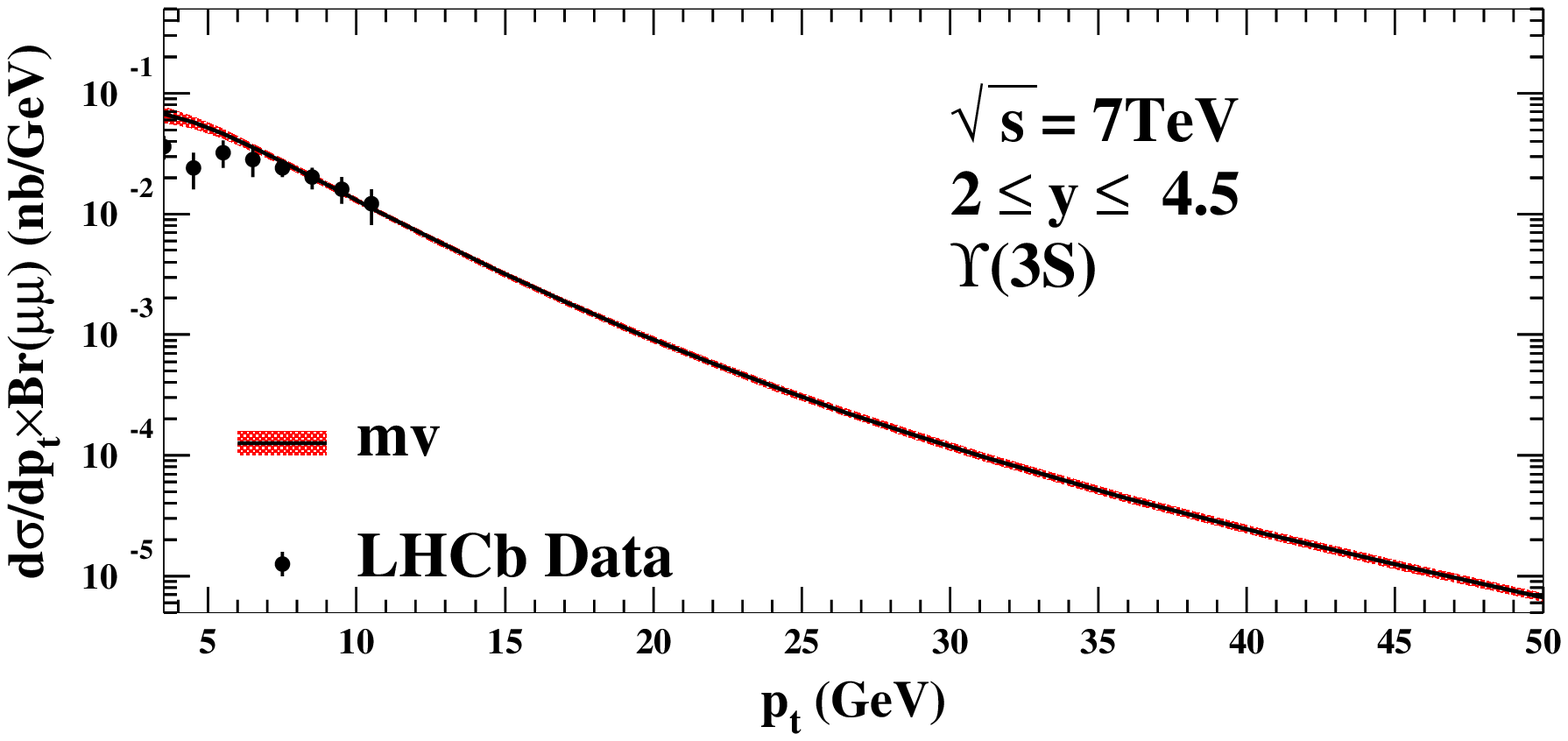} \\
  \includegraphics[width=5.0cm]{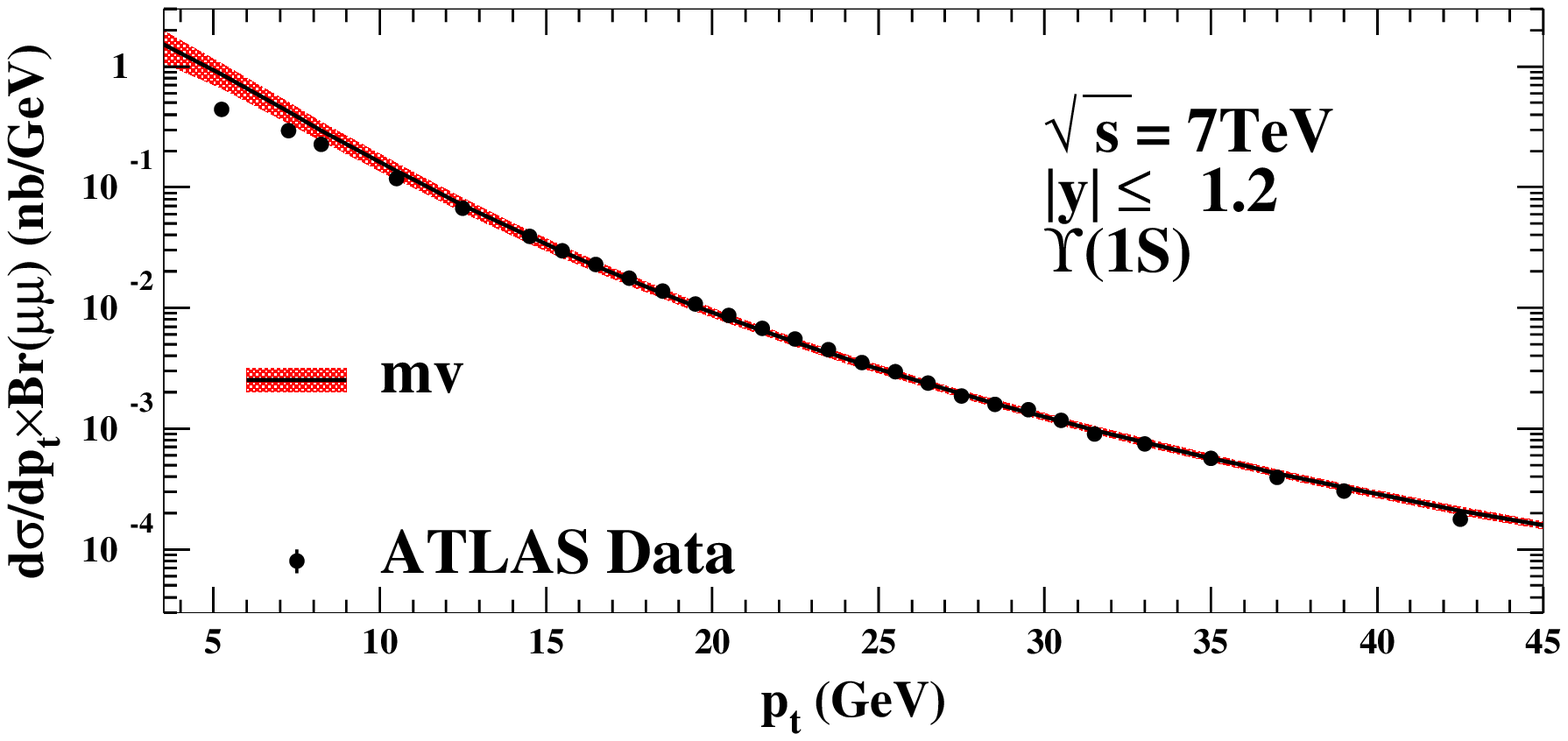}  \includegraphics[width=5.0cm]{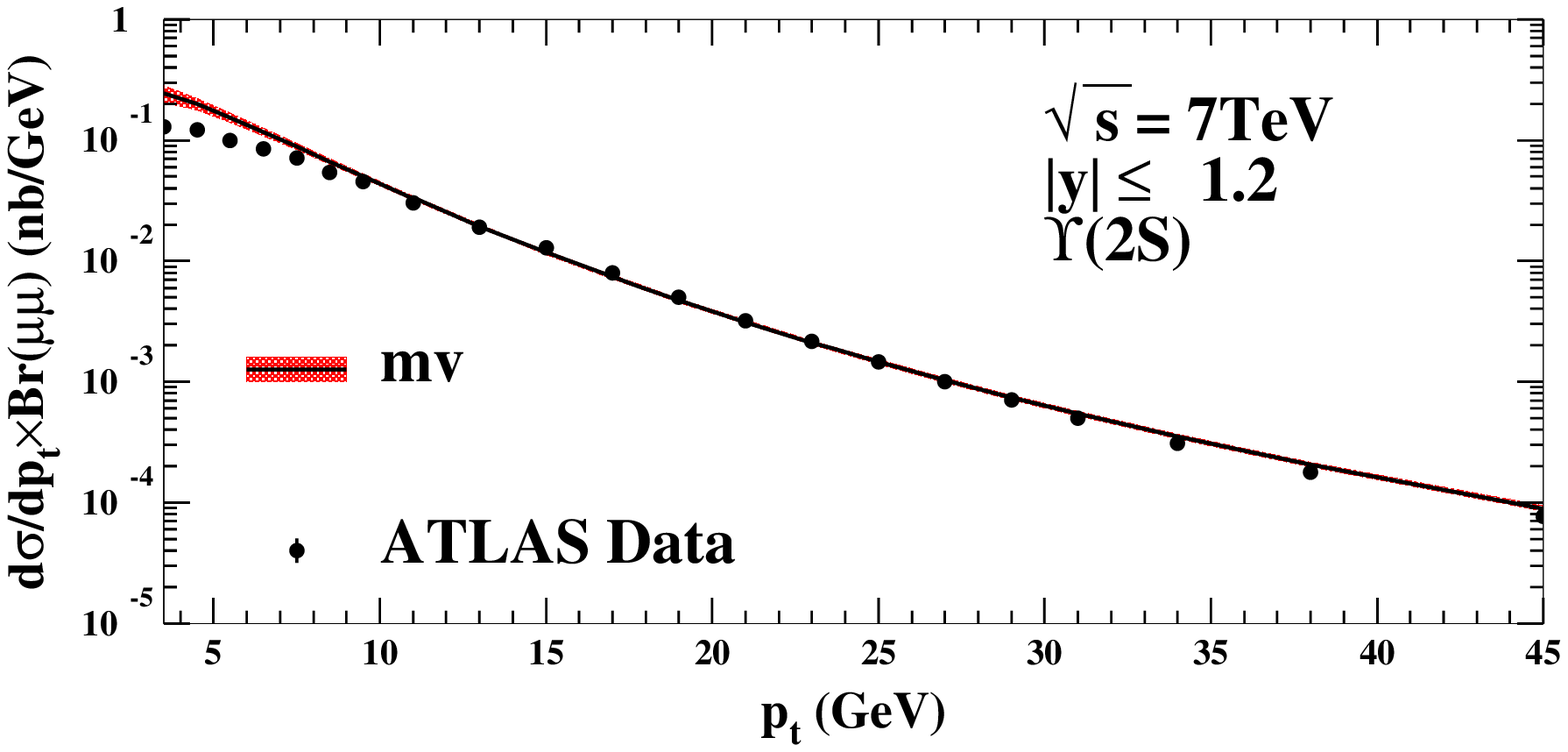} \includegraphics[width=5.0cm]{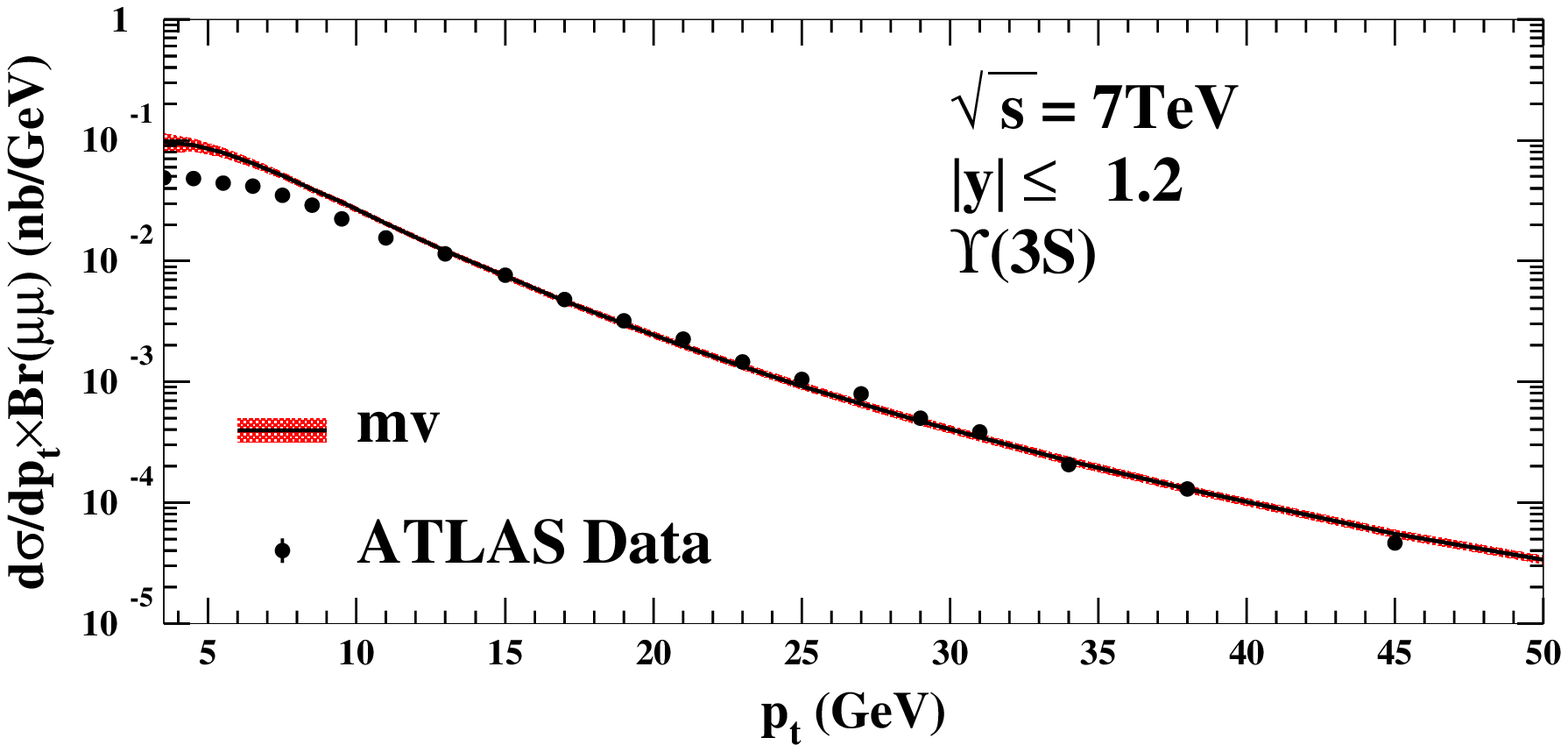}  \\
  \caption{Differential cross section for $\Upsilon$ hadroproduction at the Tevatron and LHC. From left to right:
  $\Upsilon(1S)$, $\Upsilon(2S)$ and $\Upsilon(3S)$. Rows from top to bottom correspond to different experimental conditions
  of CDF run I, CMS, LHCb, and ATLAS. The experimental data are collected from
   Refs.~\cite{Acosta:2001gv,LHCb:2012aa,Khachatryan:2010zg,Aad:2012dlq}.}
  \label{fig:dsigma-def}
\end{figure}

\begin{figure*} 
  \begin{center}
  \includegraphics[width=5.0cm]{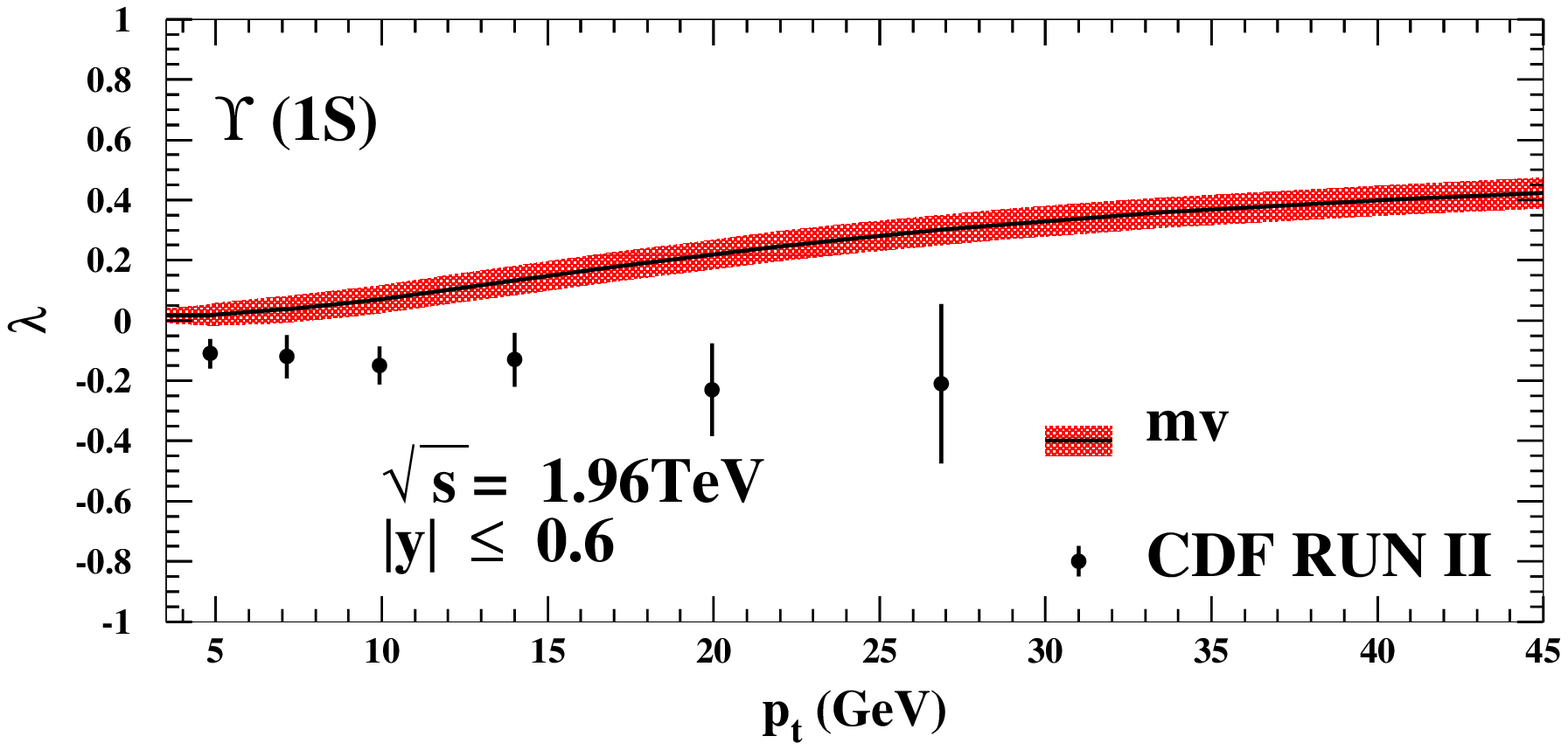} \includegraphics[width=5.0cm]{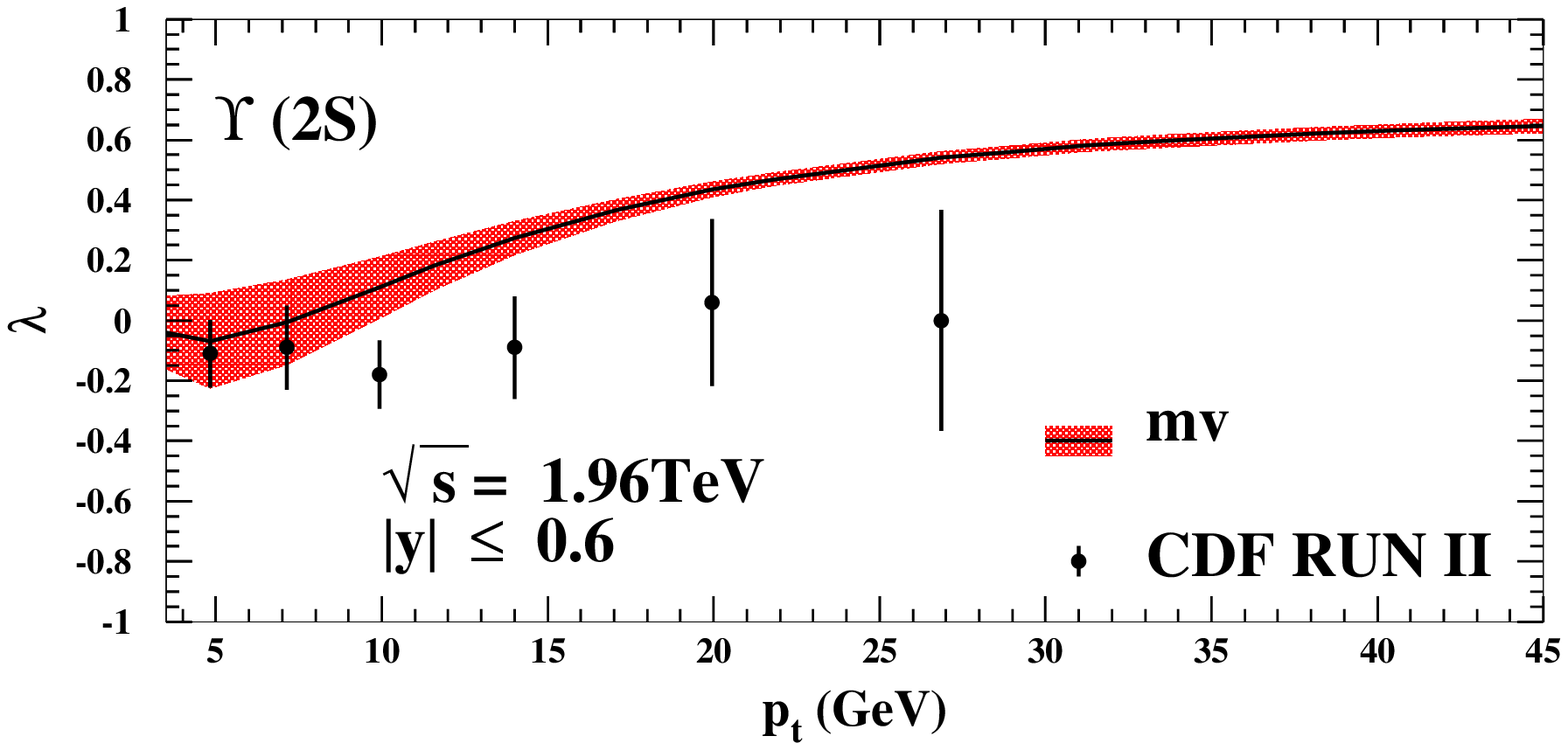} \includegraphics[width=5.0cm]{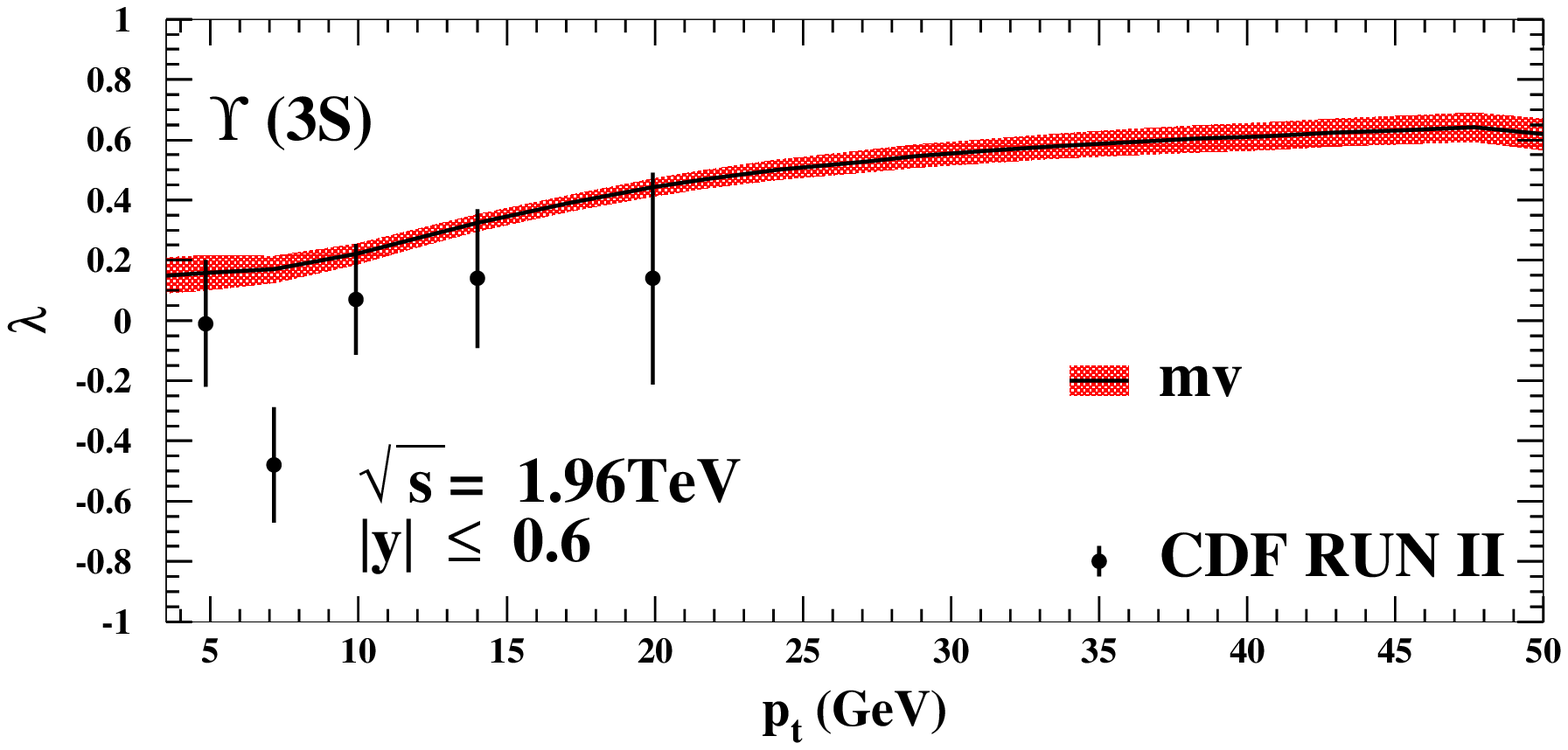} \\
  \includegraphics[width=5.0cm]{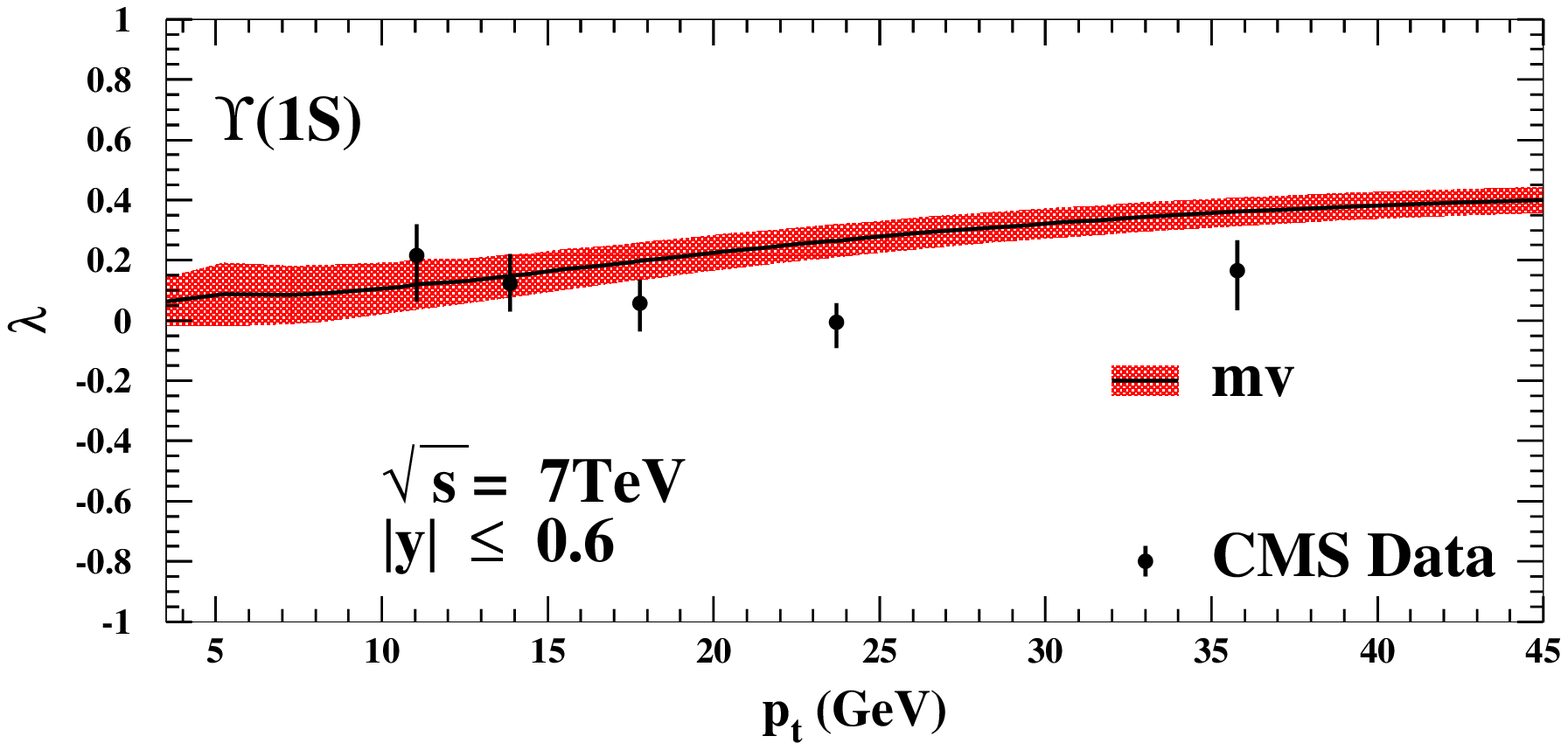}  \includegraphics[width=5.0cm]{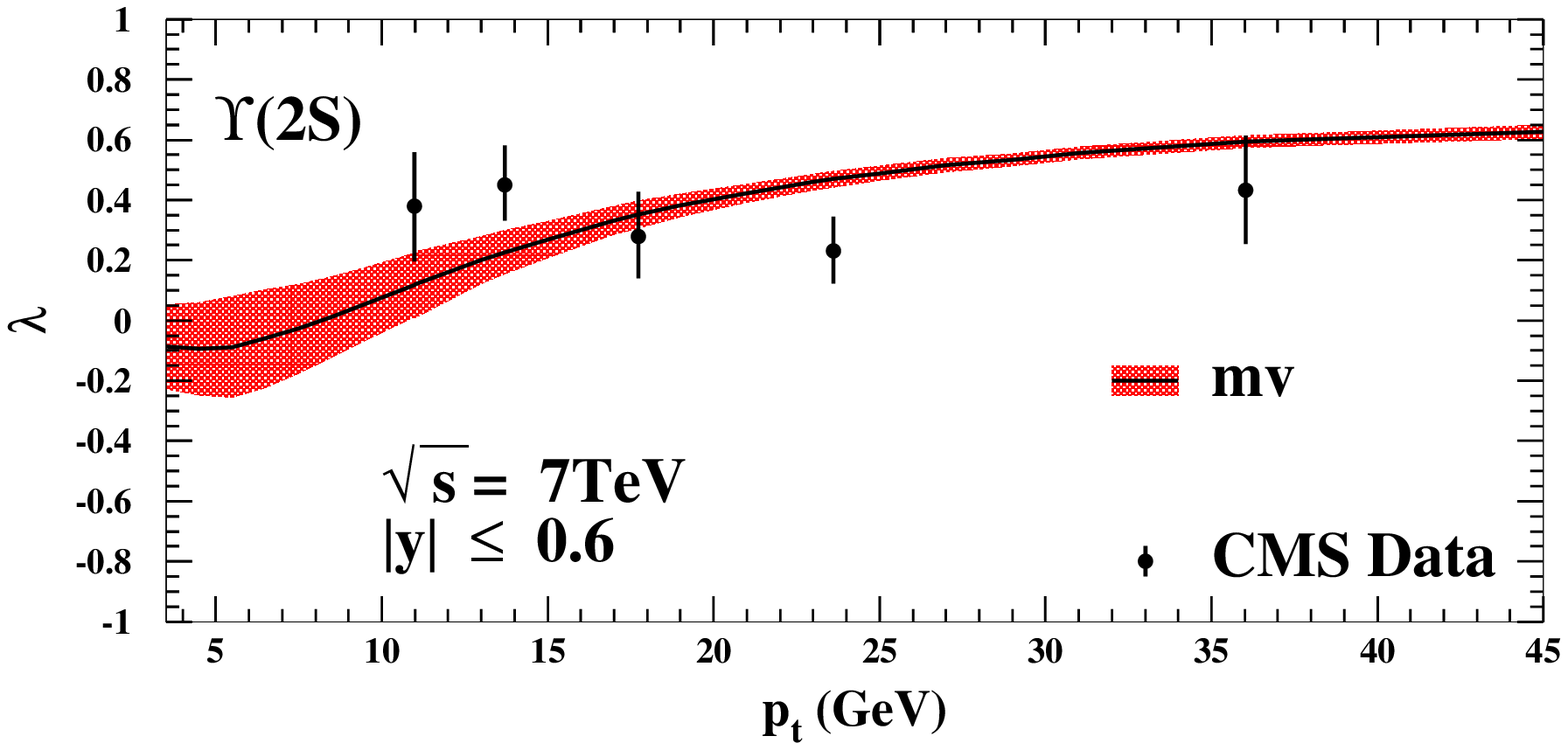}  \includegraphics[width=5.0cm]{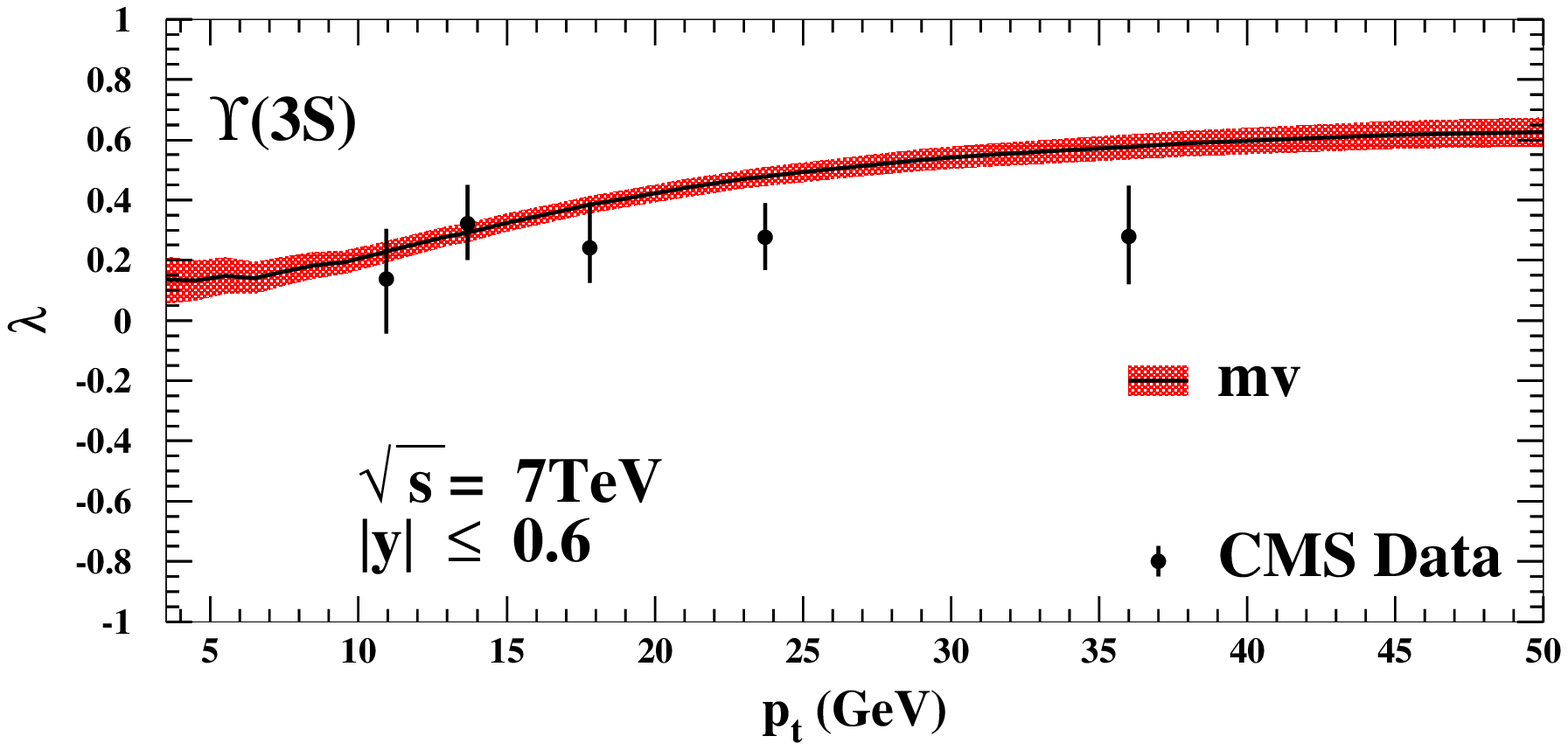} \\
  \includegraphics[width=5.0cm]{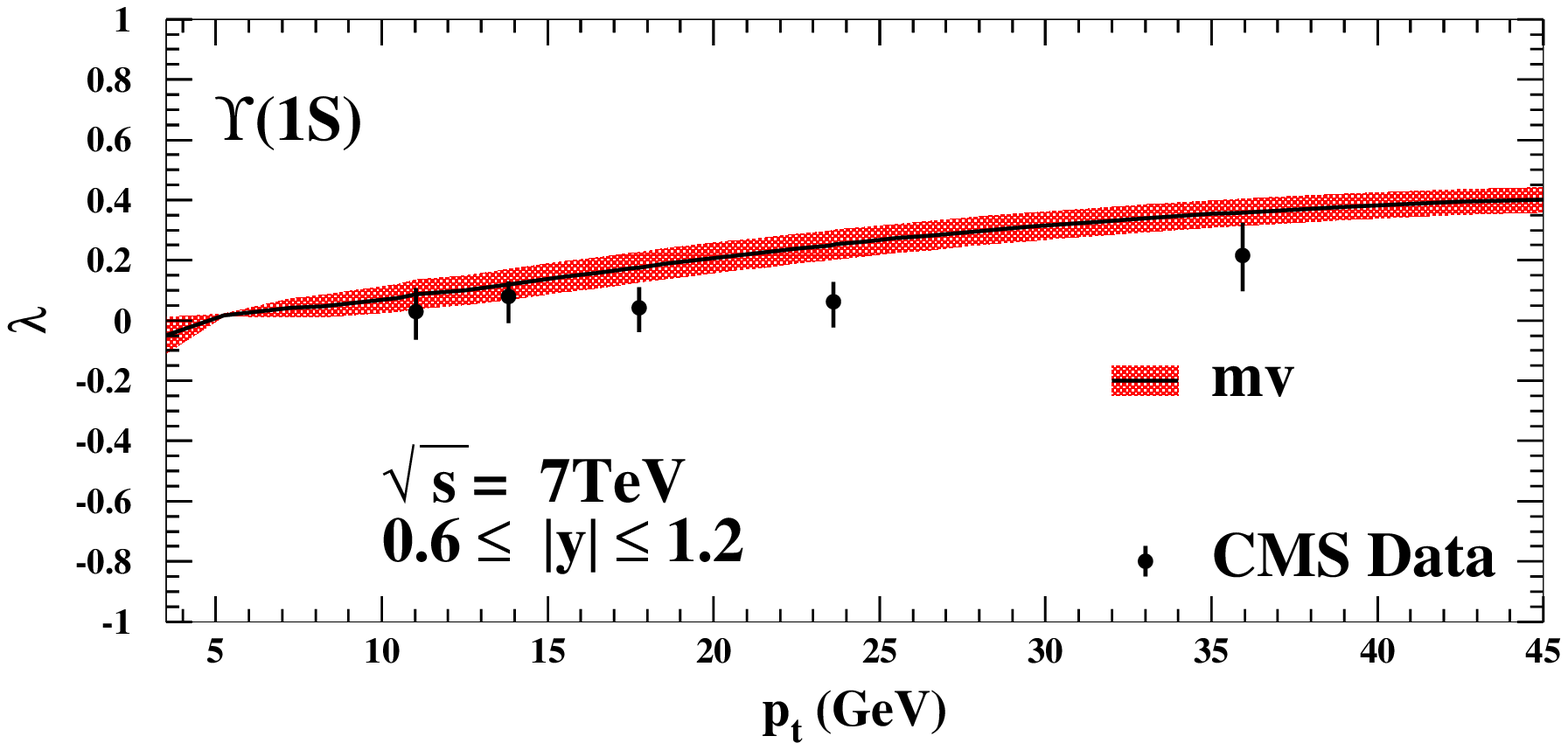}  \includegraphics[width=5.0cm]{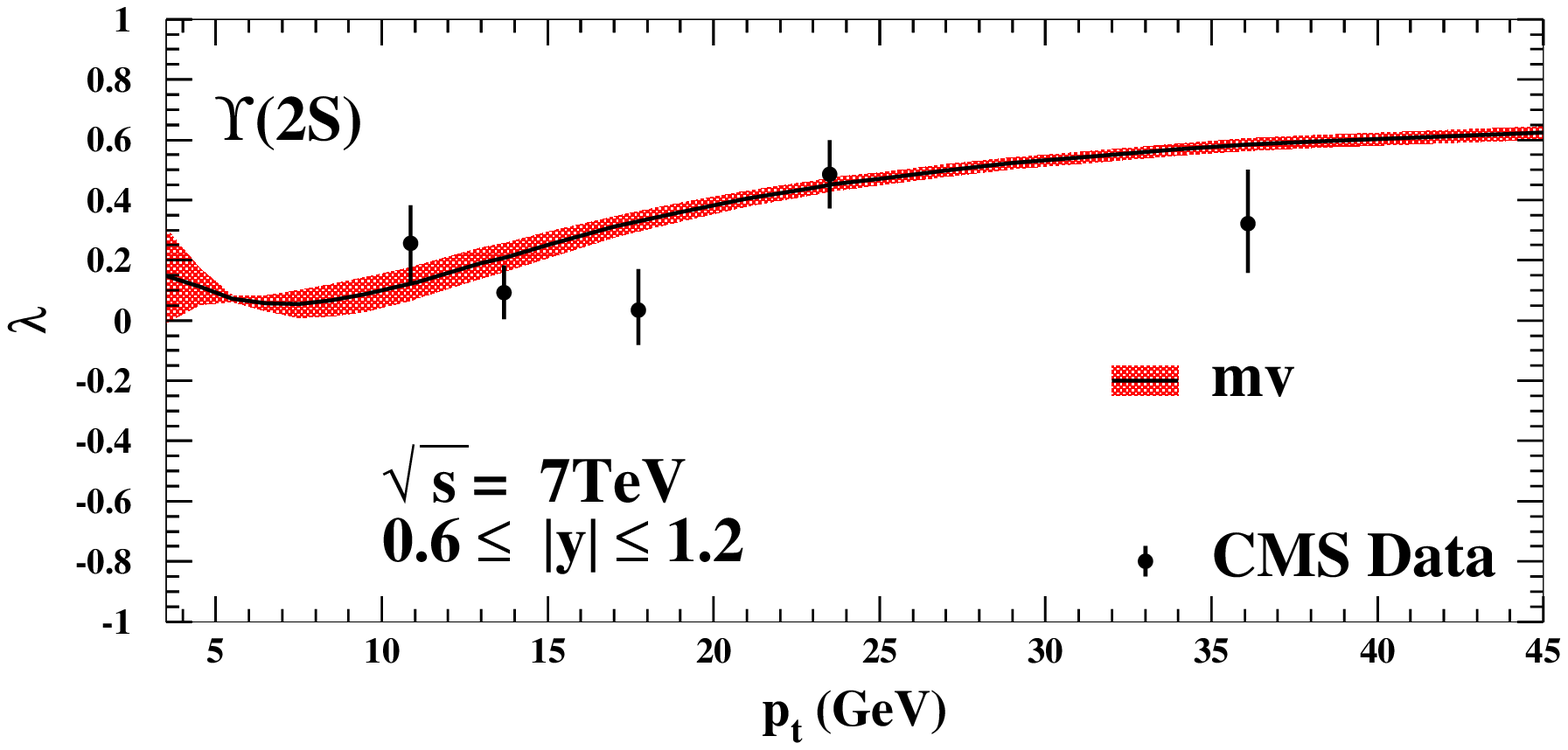}  \includegraphics[width=5.0cm]{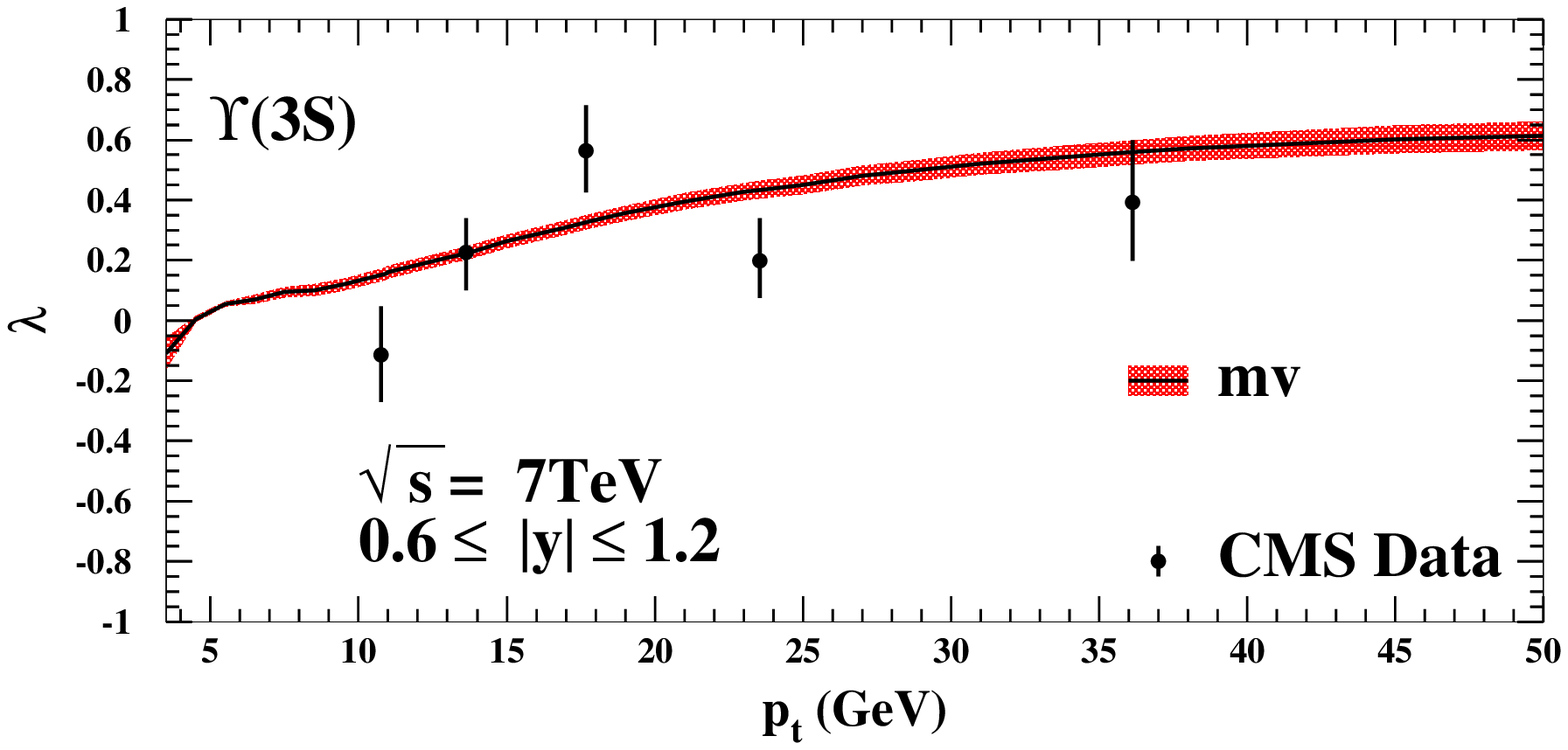} \\
  \end{center}
  \caption{Polarization parameter $\lambda$ for $\Upsilon$ hadroproduction at the Tevatron and LHC.
  From left to right: $\Upsilon(1S)$, $\Upsilon(2S)$ and $\Upsilon(3S)$. Rows from top to bottom correspond
  to different experimental conditions of CDF run II, CMS($|y|<0.6$), and CMS(0.6$<|y|<$1.2).
  The experimental data are taken from Refs.\cite{CDF:2011ag,Chatrchyan:2012woa}.}
  \label{fig:lamda-def}
\end{figure*}

The results for differential cross section of $\Upsilon$ hadroproduction are shown in Fig.~\ref{fig:dsigma-def},
while those for polarization are shown in Fig.~\ref{fig:lamda-def}. The uncertainty bands in the figures
are from the error of CO LDMEs. The experimental data are stamped with different energy $\sqrt{s}$,
rapidity cut and corresponding collaboration in all pictures.

We can see from the Fig.~\ref{fig:dsigma-def} that the results on the yield of $\Upsilon(nS)$ hadroproduction
fit all the experimental measurements very well in a wide $p_t$ range and the uncertainties are very small.
In contrast with our previous
fit results~\cite{Gong:2013qka} without $\chi_b(3P)$ feed-down contributions, where the polarization of $\Upsilon(3S)$ shows a weird steep
behaviour and can not explain the experiment measurement.
Our updated fits for $\Upsilon(3S)$ has changed a lot with the contribution of $\chi_b(3P)$ feed-down for the polarization shown in
Fig.~\ref{fig:lamda-def}, and it fits the experimental measurements very well.
For $\Upsilon(2S)$, the production is dominated by $^3S_1^{[8]}$ channel, which lead to
a slightly transverse behaviour($\lambda\approx0.6$) at high $p_t$ region.
Nevertheless, the results for polarization can explain the CMS data well,
but the distance from the CDF experimental data is still there.
This is also the case for $\Upsilon(1S)$ polarization,
although it is almost unpolarized in all $p_t$ region ($0\leq\lambda\leq0.4$).

\begin{figure*}[!ht]
  \centering
  \includegraphics[width=6.0cm]{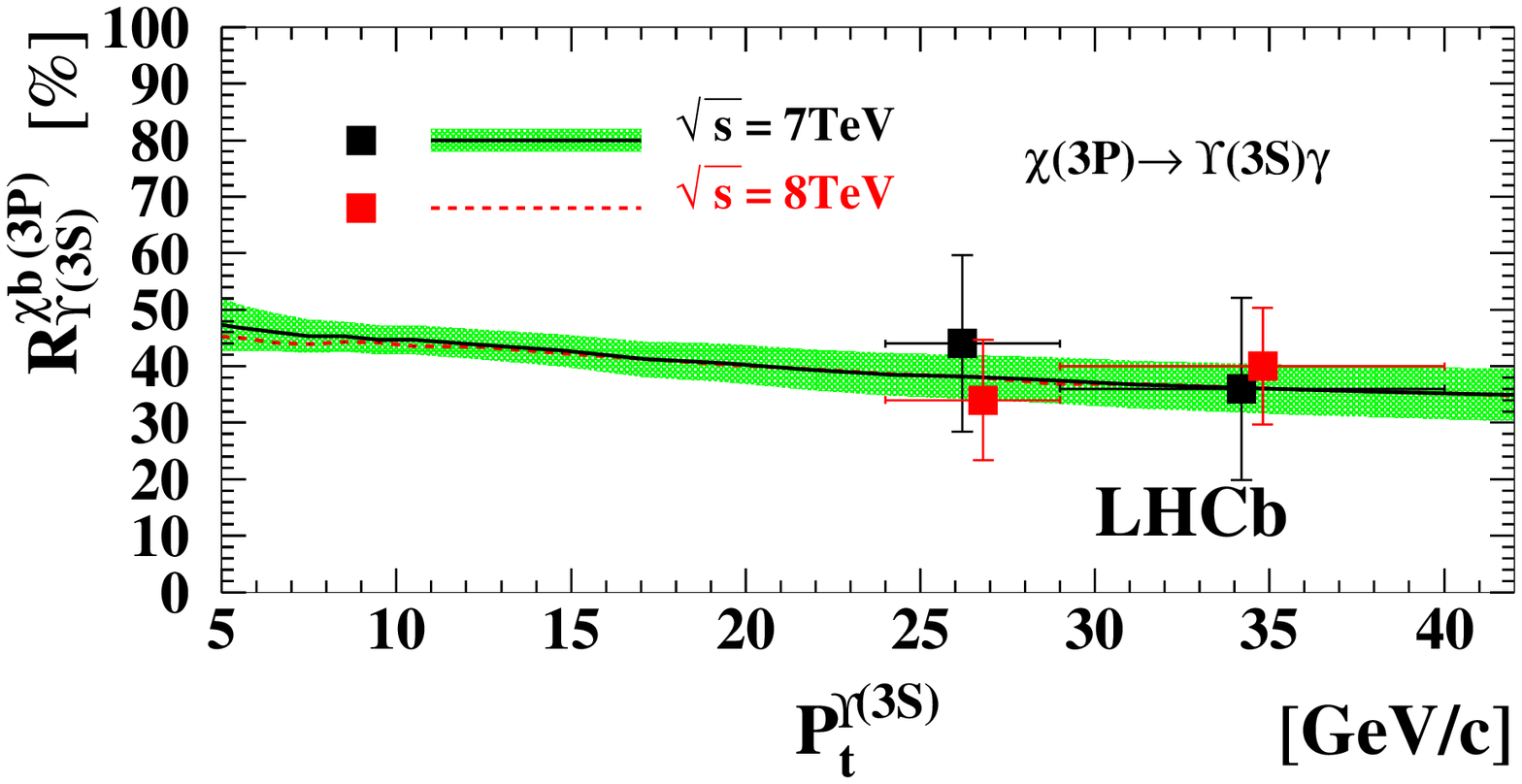}  \includegraphics[width=6.0cm]{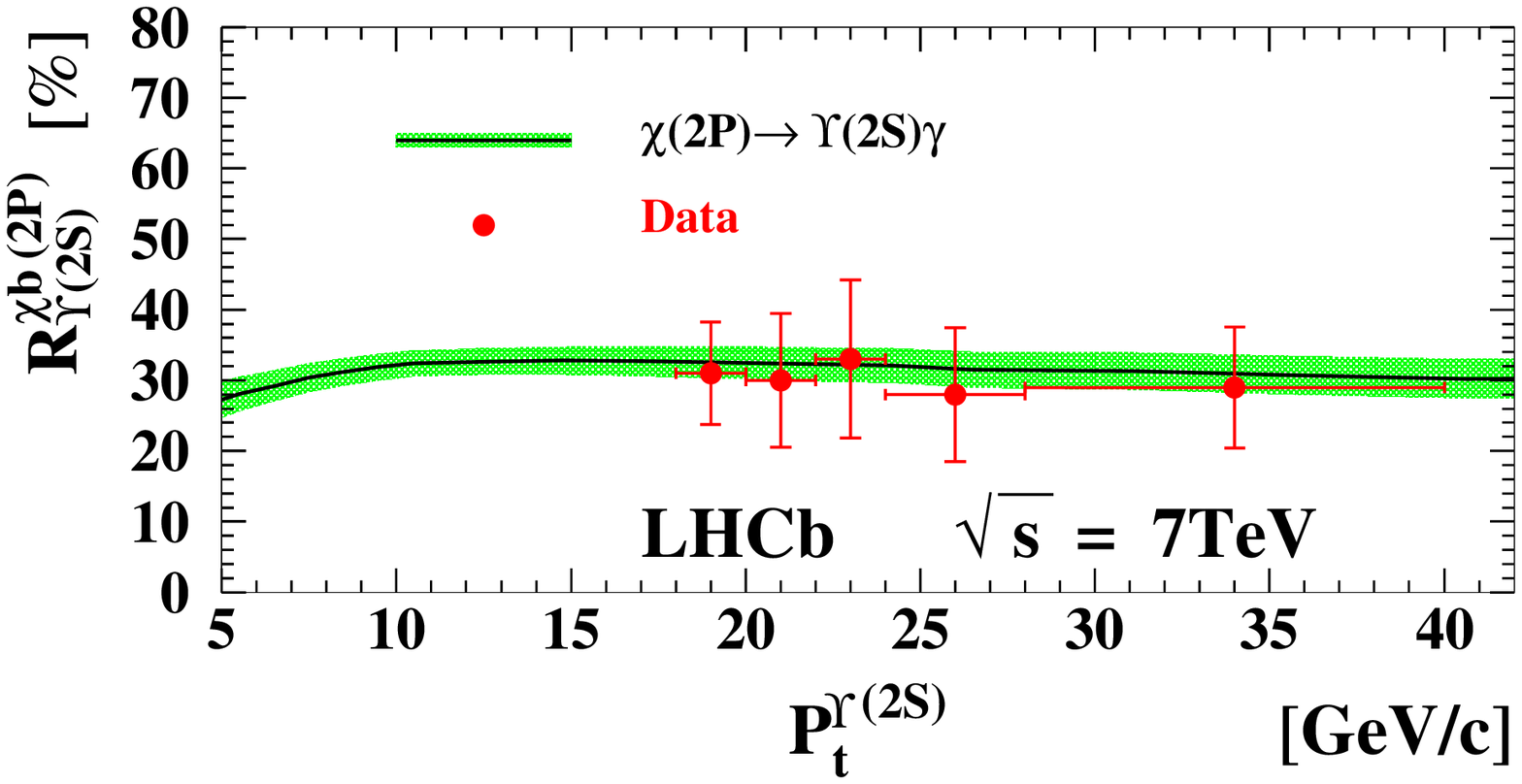} \\
  \includegraphics[width=6.0cm]{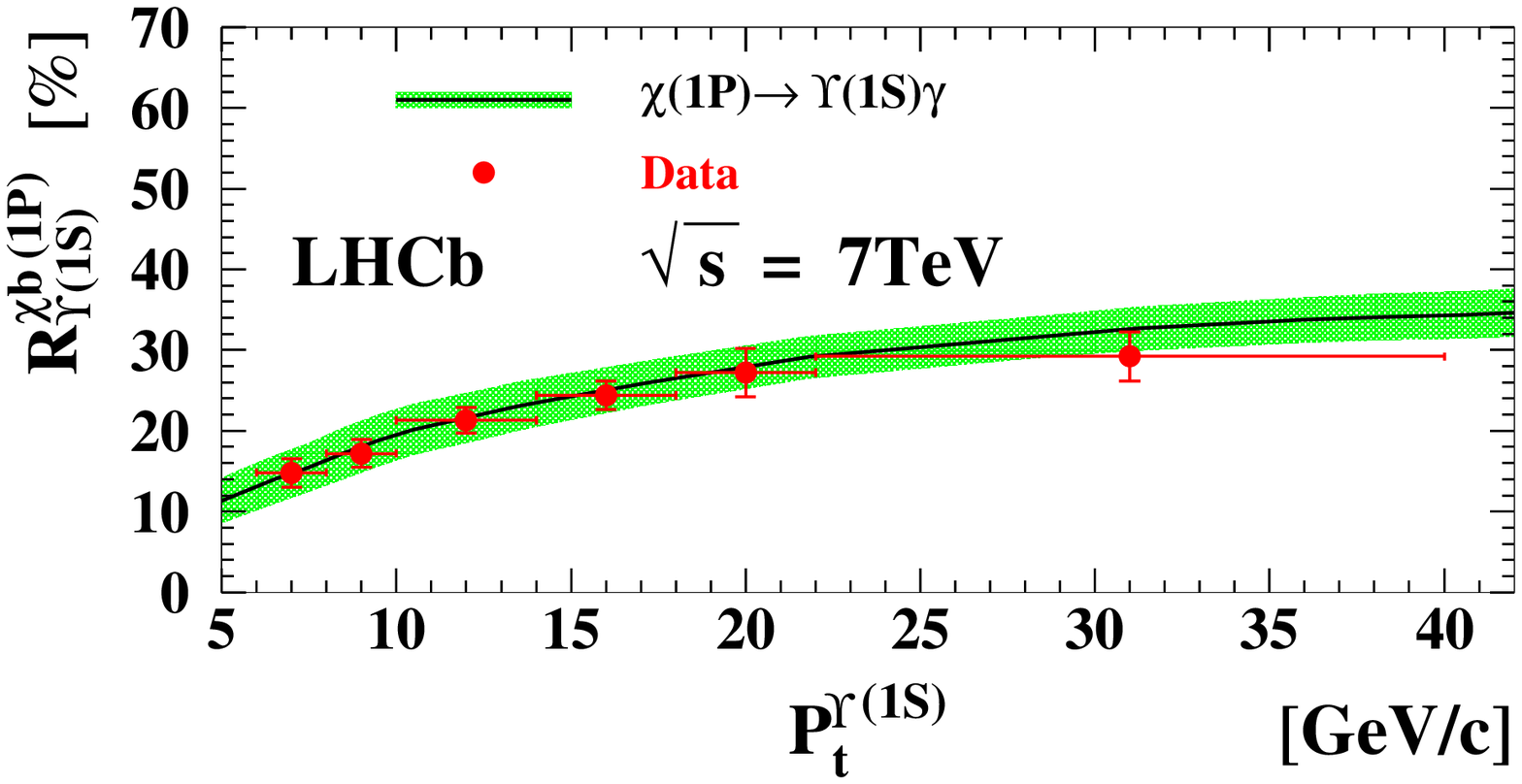} \includegraphics[width=6.0cm]{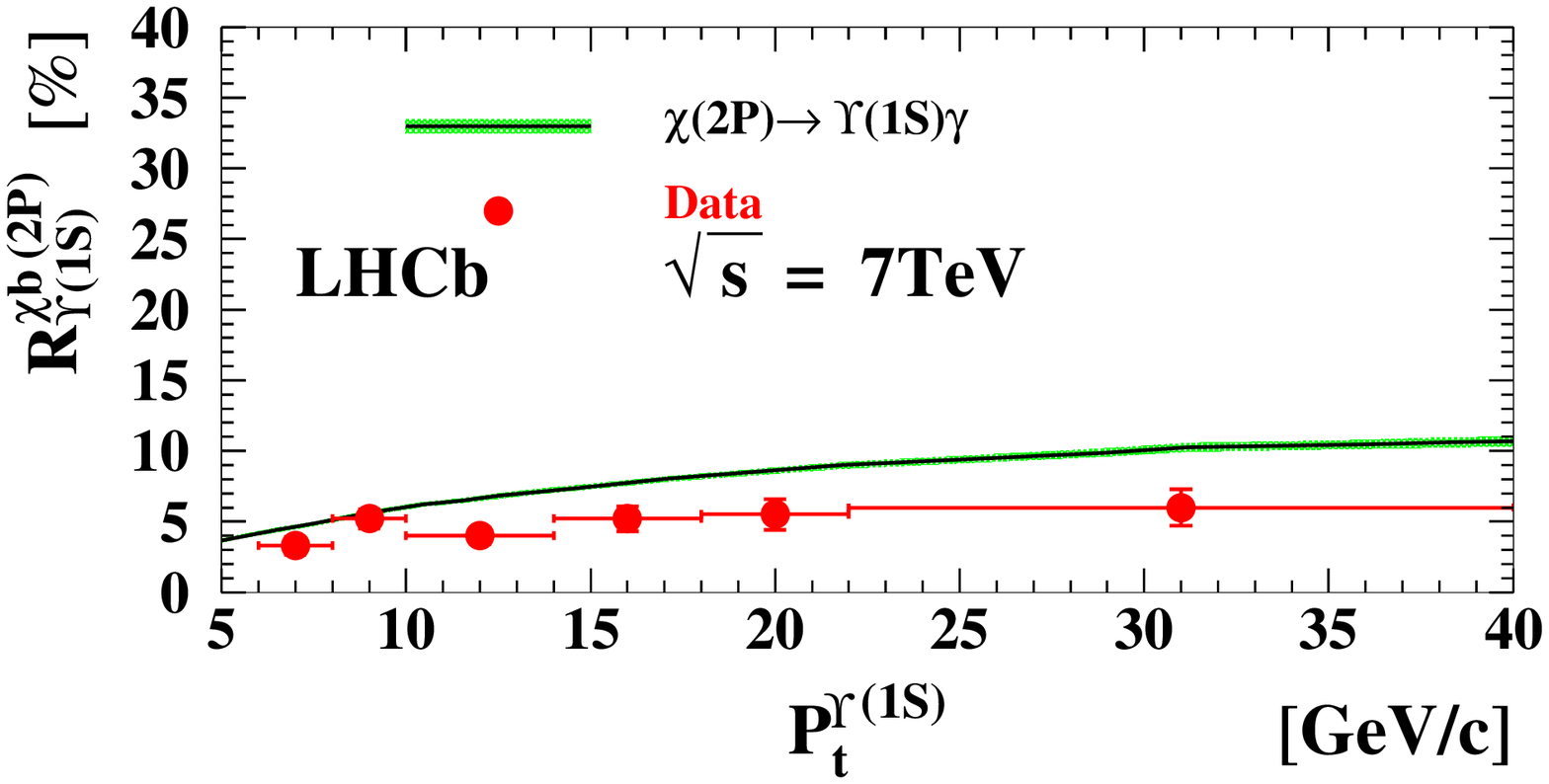} \\
  \caption{Fractions ${\cal R}^{\chi_b(mP)}_{\Upsilon(nS)}$ as functions of $p_t^{\Upsilon}$.
  From left to right: ${\cal R}^{\chi_b(3P)}_{\Upsilon(3S)}$, ${\cal R}^{\chi_b(2P)}_{\Upsilon(2S)}$ in the first row and ${\cal R}^{\chi_b(1P)}_{\Upsilon(1S)}$,${\cal R}^{\chi_b(2P)}_{\Upsilon(1S)}$ in the second row.
  The experimental data are collected from Ref.~\cite{Aaij:2014caa}. }
  \label{fig:frac-def}
\end{figure*}

\subsection{Fractions}

In Fig.~\ref{fig:frac-def}, we give the results on the fraction of $\chi_b(mP)$ feed-down to $\Upsilon(nS)$.
Only four fractions ${\cal R}^{\chi_b(3P)}_{\Upsilon(3S)}$, ${\cal R}^{\chi_b(2P)}_{\Upsilon(2S)}$,
${\cal R}^{\chi_b(1P)}_{\Upsilon(1S)}$,${\cal R}^{\chi_b(2P)}_{\Upsilon(1S)}$
are presented since in our approximation the $\chi_b(3P)$ feed-down contributions are ignored for $\Upsilon(1S,2S)$
production. For ${\cal R}^{\chi_b(3P)}_{\Upsilon(3S)}$, we give the results at $\sqrt{s}$=7 TeV (black solid line)
and $\sqrt{s}=$8 TeV (red dotted line), but they are almost overlapping.
For the other three cases, the results at $\sqrt{s}=$8 TeV are not included since in the experimental
measurements they are almost the same with the $\sqrt{s}=$7 TeV~\cite{Aaij:2014caa} and it is believed that theoretical
predictions for the fractions is almost unchanged for the centre-of-mass energy $\sqrt{s}=$7,8 TeV.
Our results for ${\cal R}^{\chi_b(nP)}_{\Upsilon(nS)}$ show different behaviours for each $n=1,2,3$, they are
increasing, flat ($p_t>$10 GeV), slightly decreasing respectively as $p_t$ increases. And
all of them fit the experimental data very well. The remaining fraction ${\cal R}^{\chi_b(2P)}_{\Upsilon(1S)}$ shows a
increasing behaviour but the theoretical result overshoot the data about factor 2.

\subsection{Ratio of $\sigma(\chi_{b2}(1P))/\sigma(\chi_{b1}(1P))$}
Recently, new measurements on the ration of cross sections of $\sigma[\chi_{b2}(1P)]$ to $\sigma[\chi_{b1}(1P)]$
are reported by LHCb~\cite{Aaij:2014hla} and CMS collaborations~\cite{Khachatryan:2014ofa}.

\begin{figure}[!ht]
  \begin{center}
  \includegraphics[width=6.0cm]{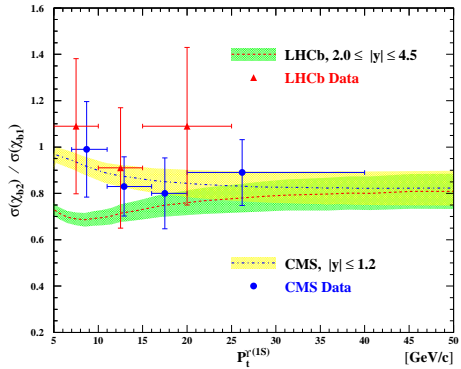} \\
  \caption{The ratio of the cross sections of $\chi_{b2}(1P)$ to $\chi_{b1}(1P)$ production , as a function of $p_t$.
  The yellow band is the results for CMS experimental condition
  while the green band is for LHCb experimental condition.
  The data are taken from Ref.~\cite{Aaij:2014hla}~\cite{Khachatryan:2014ofa}.}
  \label{fig:frac-b2b1-def}
  \end{center}
\end{figure}

As a way to check our fitting results for the LDMEs, we present the prediction of
the cross section ratio $\sigma[\chi_{b2}(1P)]/\sigma[\chi_{b1}(1P)]$ in Fig.~\ref{fig:frac-b2b1-def},
where the green band is the results for LHCb experimental condition while the yellow band is for CMS condition.
As the picture shows that the CMS experimental data can be explained by the theory,
while the LHCb data is underestimated. One may notice that here is an approximation for the CMS theoretical prediction
that the rapidity range is $|y|<1.2$, while for the experimental data in Ref.~\cite{Khachatryan:2014ofa} the rapidity range is $|y|<1.5$.
Although the prediction condition is not exactly the same with data, we think the approximation is fairly enough.

\section{More fitting results}\label{charp:4}

While we are preparing this work, a similar work was preprinted by H.Han $et~al.$~\cite{Han:2014kxa},
where they estimated the branching
ratios of $\chi_b(3P)$ feed-down to $\Upsilon(1S,2S,3S)$ and fitted the productions and polarizations.
In our fit in the former section, we ignored the $\chi_b(3P)$ feed-down contributions
for $\Upsilon(1S,2S)$ since it is really small. To give a more rigorous feed-down contributions would be instructive.
Meanwhile, the NRQCD factorization scale $\mu_{\Lambda}$ dependence dose exist at the fixed order results. To explore
the uncertainty from these two aspects,  we refit the production and polarization and obtain two set of new LDMEs
for $\mu_{\Lambda}=m_b v=1.5$ GeV and $\mu_{\Lambda}=m_b$ together with the branching
ratios ${\cal B}[\chi_{bJ}(3P)\rightarrow\Upsilon(nS)]$ estimated in Ref.~\cite{Han:2014kxa} TABLE~I,
and the numerical results are presented in
Table.~\ref{tab:LDMEs-mv} and Table.~\ref{tab:LDMEs-mb} respectively.

For convenience, we define three short statements,
\begin{itemize}
  \item \textit{default scheme} ($\mu_{\Lambda}=m_b v=1.5$ GeV and our naive estimation for branching ratios
  ${\cal B}[\chi_b(3P)\rightarrow \Upsilon(3S)]={\cal B}[\chi_b(2P)\rightarrow \Upsilon(2S)]$,
  ${\cal B}[\chi_b(3P)\rightarrow \Upsilon(1S,2S)]=0$),

  \item \textit{Han's scheme} ($\mu_{\Lambda}=m_b$ and branching ratios
  ${\cal B}[\chi_{bJ}(3P) \rightarrow\Upsilon(nS)]$ estimated in Ref.~\cite{Han:2014kxa} TABLE~I),

  \item \textit{mix scheme} ($\mu_{\Lambda}=m_b v=$1.5 GeV and branching ratios
   ${\cal B}[\chi_{bJ}(3P) \rightarrow\Upsilon(nS)]$ estimated in Ref.~\cite{Han:2014kxa} TABLE~I).
\end{itemize}

For the mix scheme in Table.~\ref{tab:LDMEs-mv}, some changes compared with default scheme in Table.~\ref{tab:LDMEs-def} should
be mentioned.  For $\Upsilon(3S)$, the dominate CO channel becomes $^3S_1^{[8]}$ and the numerical value for
$\langle{\cal O}^{\chi_b(3P)}(^{3}S^{[8]}_{1})\rangle$ becomes larger.
This is not at all a surprise since we know the branching ratios ${\cal B}[\chi_b(3P)\rightarrow\Upsilon(3S)]$
used in the default scheme is much larger than it is in the mix scheme. This difference is consistent with
the change of the proportions for LDMEs in two cases. For $\Upsilon(2S)$ and $\Upsilon(1S)$,
the LDMEs values have smaller changes.

In comparison with the mix scheme, for Han's scheme in Table.~\ref{tab:LDMEs-mb},
the LDMEs for $\Upsilon(2S)$ changed a lot because the $^1S_0^{[8]}$ channel becomes
negative and the $^3P_J^{[8]}$ channel gives positive contributions. Besides, the value of
$\langle{\cal O}^{\chi_b(1P)}(^{3}S^{[8]}_{1})\rangle$ becomes larger,
which may impact the behaviour of the ratio $\sigma[\chi_{b2}(1P)]/\sigma[\chi_{b1}(1P)]$.

\begin{table}[!ht]
  \begin{center}
  \caption{ \label{tab:LDMEs-mv} Same as Table.~\ref{tab:LDMEs-def}, except that we choose the branching ratio
  ${\cal B}[\chi_{bJ}(3P)\rightarrow\Upsilon(nS)]$ in Ref.~\cite{Han:2014kxa}, so-called mix scheme.}
  \footnotesize
  \begin{tabular*}{160mm}{@{\extracolsep{\fill}}cccccc}
  \hline\hline
  state ~&~ $\langle{\cal O}^{\Upsilon(nS)}(^{1}S^{[8]}_{0})\rangle$ ~&~
  $\langle{\cal O}^{\Upsilon(nS)}(^{3}S^{[8]}_{1})\rangle$ ~&~
  $\langle{\cal O}^{\Upsilon(nS)}(^{3}P^{[8]}_{0})\rangle/m_b^2$ ~&~
  state ~&~ $\langle{\cal O}^{\chi_b(mP)}(^{3}S^{[8]}_{1})\rangle$ \\ [0.1cm]
  \hline
  $\Upsilon(1S)$ ~&~ 10.1 $\pm$ 2.23 ~&~ 0.73 $\pm$ 0.22 ~&~ -0.23 $\pm$ 0.50 ~&~ $\chi_b(1P)$ ~&~ 0.91 $\pm$ 0.06  \\
  $\Upsilon(2S)$ ~&~ 1.19 $\pm$ 1.93 ~&~ 1.88 $\pm$ 0.23 ~&~ -0.01 $\pm$ 0.42 ~&~ $\chi_b(2P)$ ~&~ 1.07 $\pm$ 0.12  \\
  $\Upsilon(3S)$ ~&~ -0.15 $\pm$ 0.90 ~&~ 1.53 $\pm$ 0.12 ~&~ -0.02 $\pm$ 0.19 ~&~ $\chi_b(3P)$ ~&~ 1.76 $\pm$ 0.14 \\
  \hline\hline
  \end{tabular*}
  \end{center}
\end{table}

\begin{table}[!ht]
  \begin{center}
  \caption{ \label{tab:LDMEs-mb}Same as Table.~\ref{tab:LDMEs-mv}, except that we choose $\mu_{\Lambda}=m_b$, the Han's scheme.}
  \footnotesize
  \begin{tabular*}{160mm}{@{\extracolsep{\fill}}cccccc}
  \hline\hline
  state ~&~ $\langle{\cal O}^{\Upsilon(nS)}(^{1}S^{[8]}_{0})\rangle$ ~&~
  $\langle{\cal O}^{\Upsilon(nS)}(^{3}S^{[8]}_{1})\rangle$ ~&~
  $\langle{\cal O}^{\Upsilon(nS)}(^{3}P^{[8]}_{0})\rangle/m_b^2$ ~&~
  state ~&~ $\langle{\cal O}^{\chi_b(mP)}(^{3}S^{[8]}_{1})\rangle$ \\ [0.1cm]
  \hline
  $\Upsilon(1S)$ ~&~ 11.6 $\pm$ 2.61 ~&~ 0.47 $\pm$ 0.41 ~&~ -0.49 $\pm$ 0.59 ~&~ $\chi_b(1P)$ ~&~ 1.16 $\pm$ 0.07  \\
  $\Upsilon(2S)$ ~&~ -0.59 $\pm$ 2.31 ~&~ 2.94 $\pm$ 0.40 ~&~ 0.28 $\pm$ 0.52 ~&~ $\chi_b(2P)$ ~&~ 1.50 $\pm$ 0.21  \\
  $\Upsilon(3S)$ ~&~ -0.18 $\pm$ 1.40 ~&~ 1.52 $\pm$ 0.33 ~&~ -0.01 $\pm$ 0.30 ~&~ $\chi_b(3P)$ ~&~ 1.92 $\pm$ 0.34 \\
  \hline\hline
  \end{tabular*}
  \end{center}
\end{table}

The results of the yield and polarization in both Han's scheme and the mix scheme are almost the same as those in default scheme,
and the yield results describe the experimental data very well and so as the polarization of CMS measurement,
but not the CDF measurements. We would not show these pictures here to avoid unnecessary repetition
since there is little difference compared with results in default scheme shown in
Fig.~\ref{fig:dsigma-def} (for yield) and Fig.~\ref{fig:lamda-def} (for polarization).

\begin{figure*}[!ht]
  \centering
  \includegraphics[width=5.0cm]{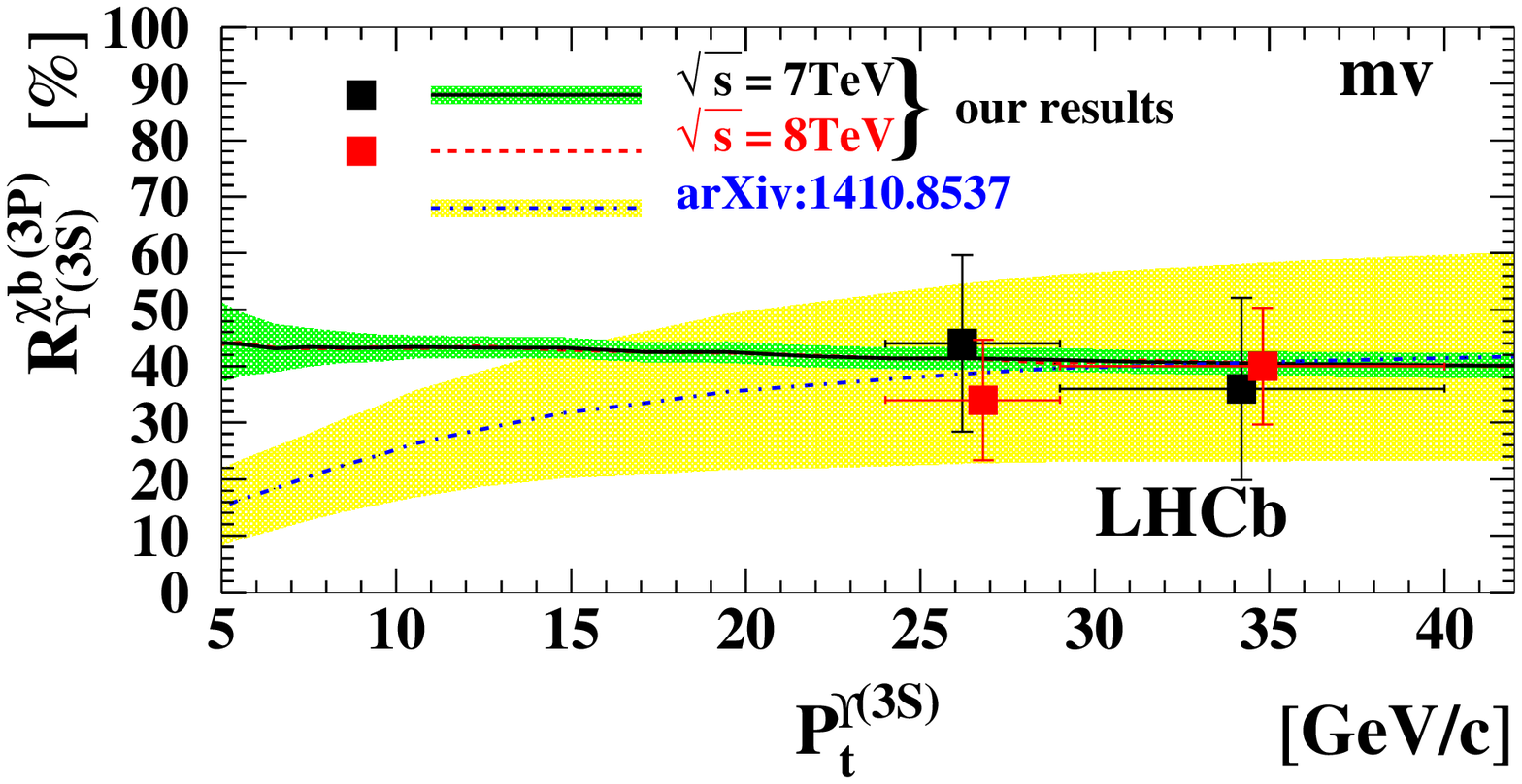}  \includegraphics[width=5.0cm]{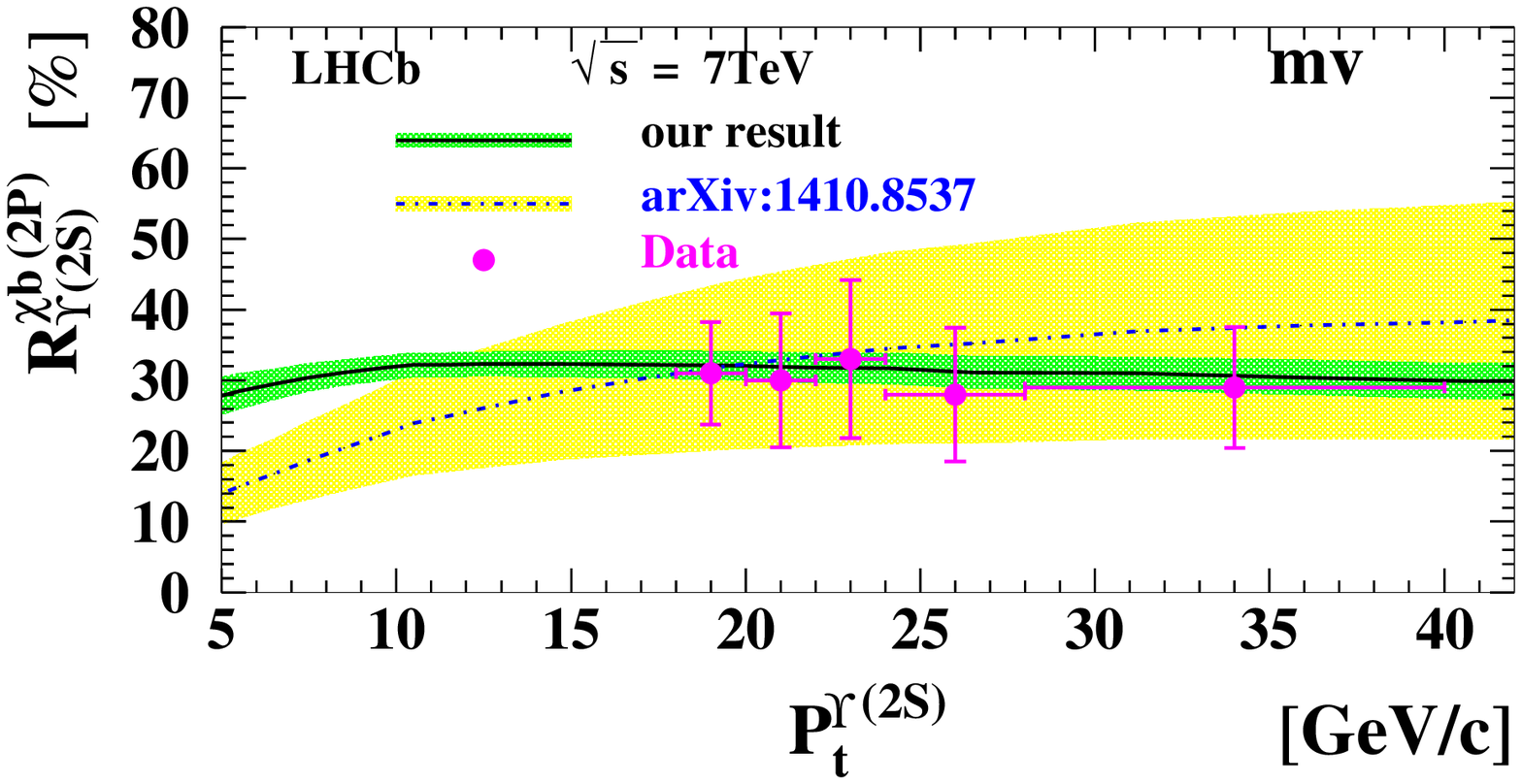}
  \includegraphics[width=5.0cm]{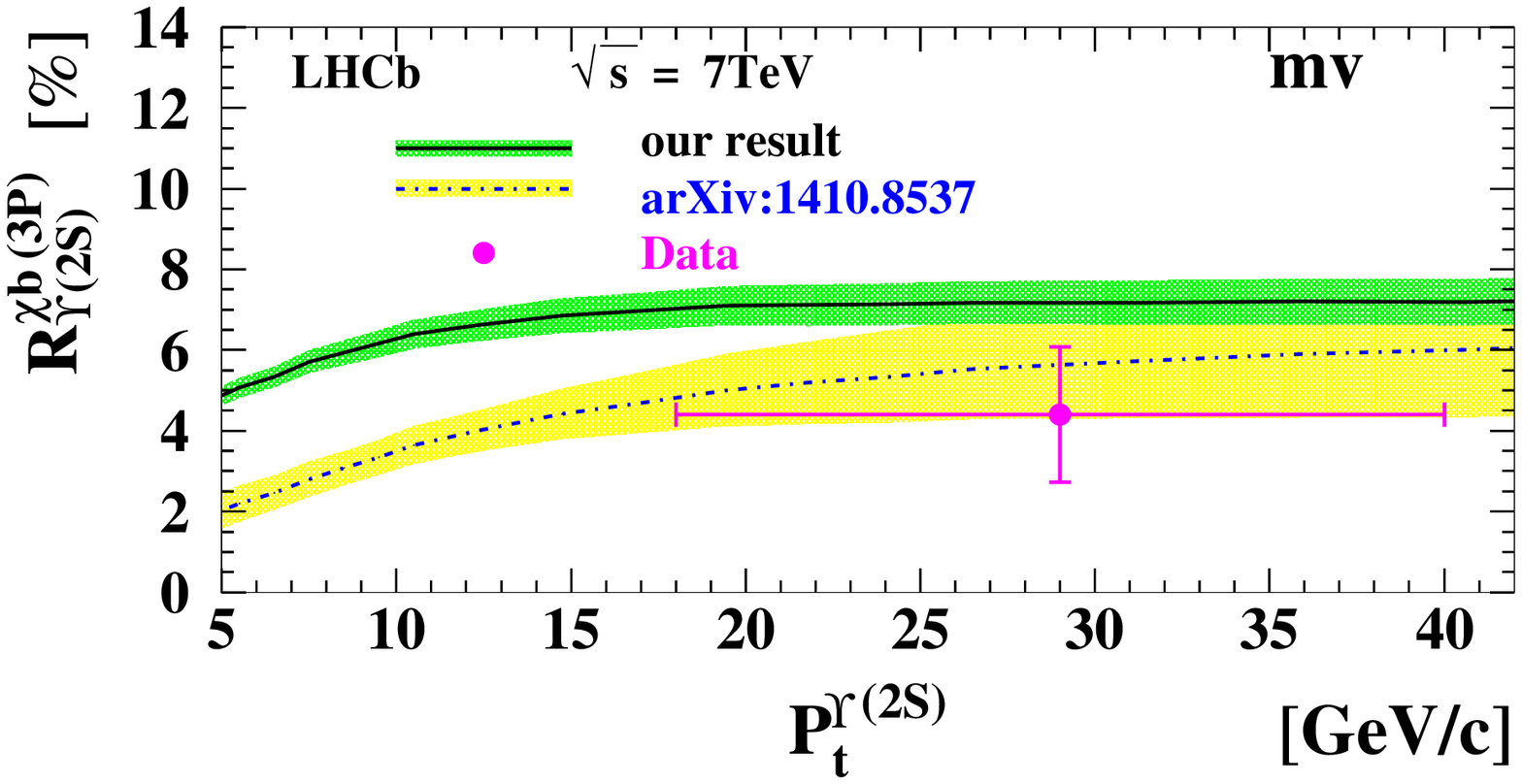} \\
  \includegraphics[width=5.0cm]{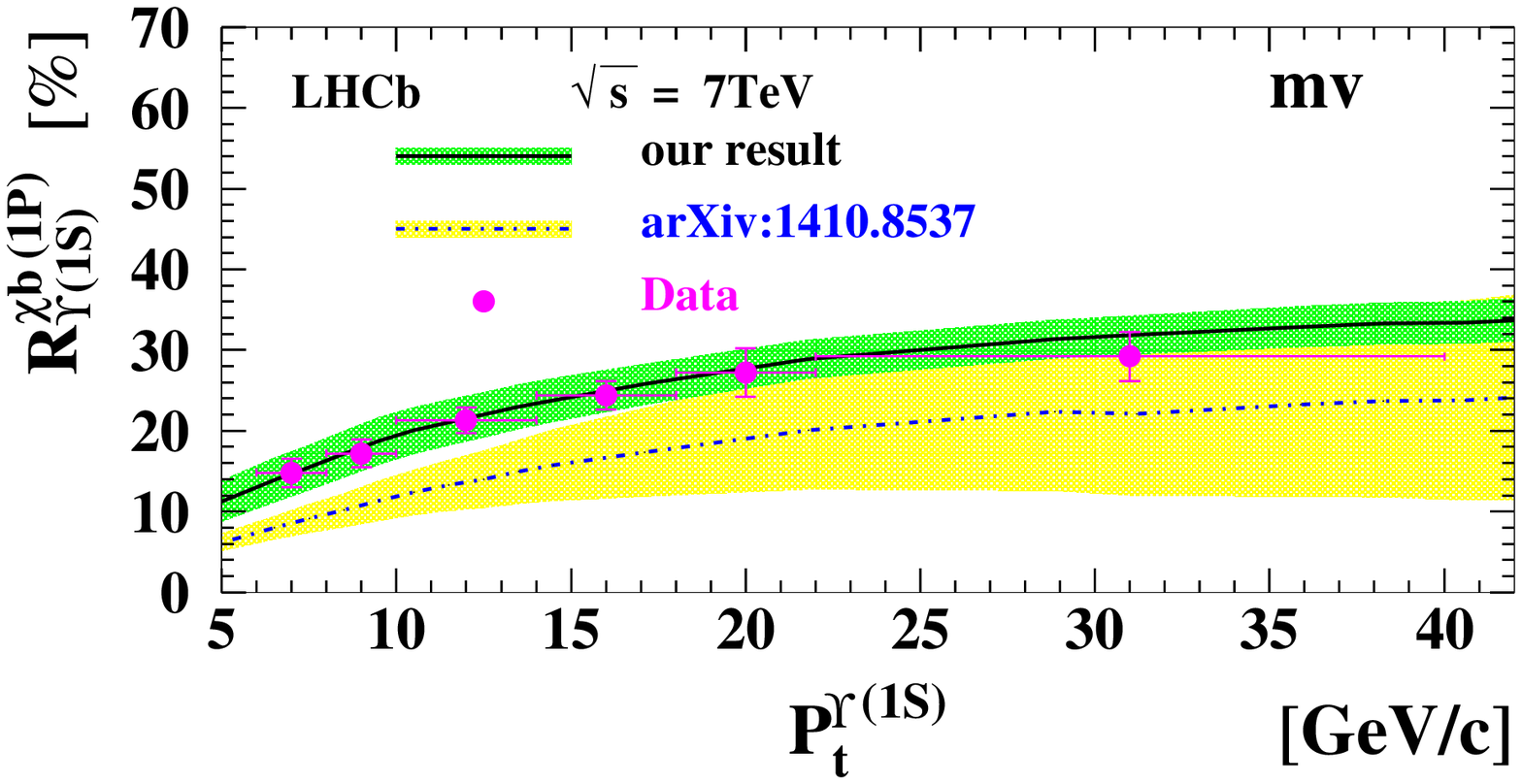} \includegraphics[width=5.0cm]{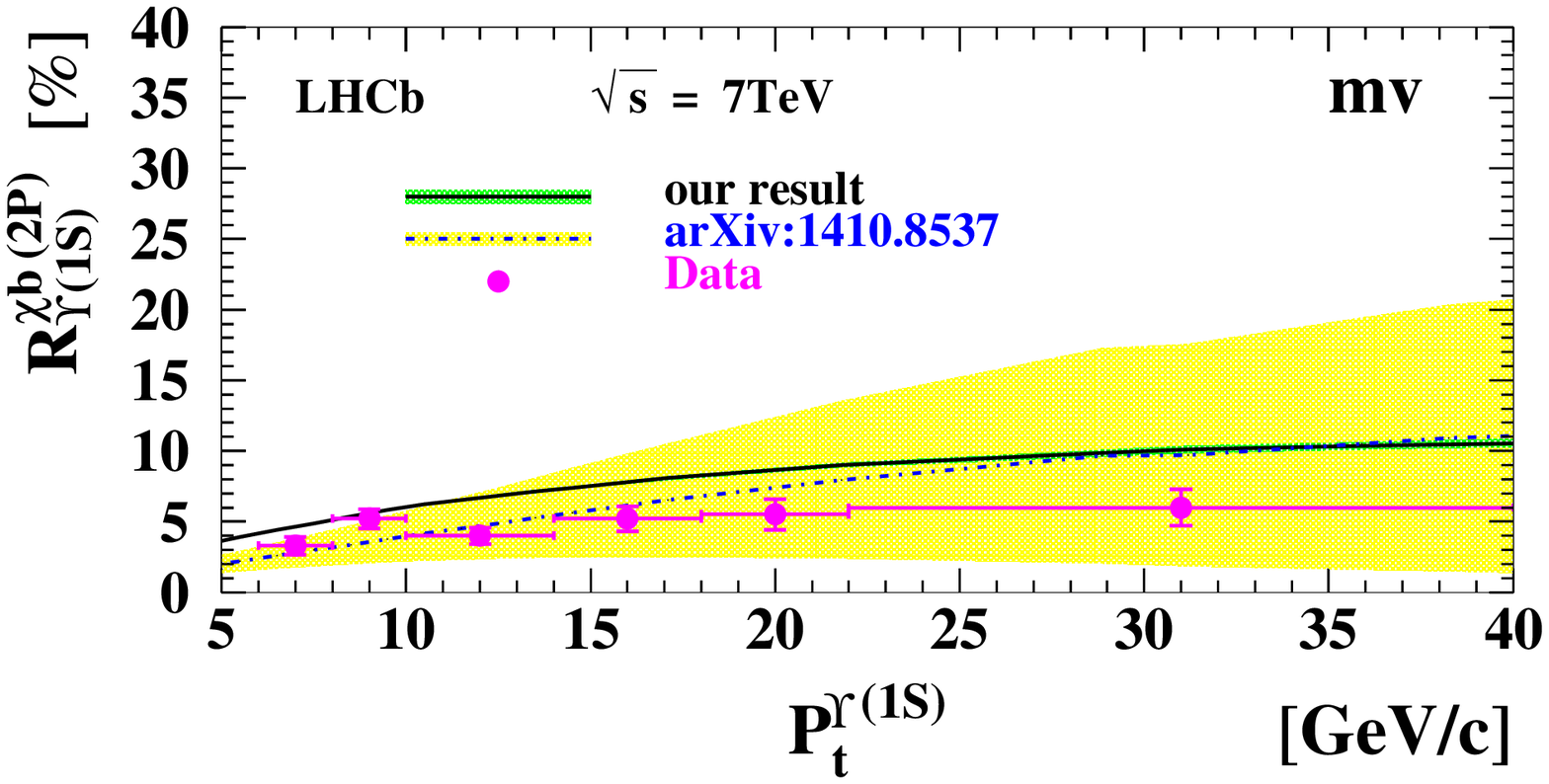}
  \includegraphics[width=5.0cm]{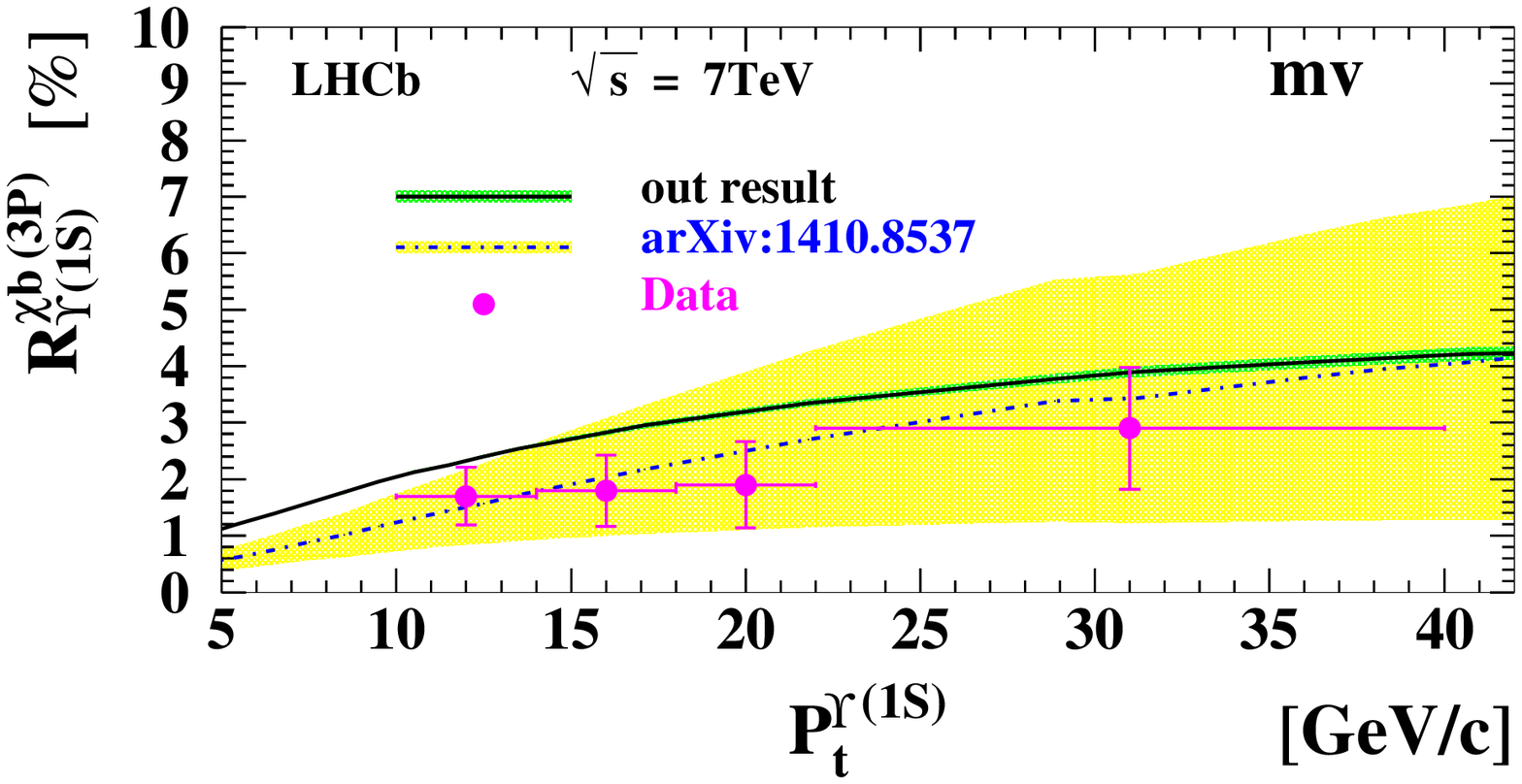} \\
  \includegraphics[width=5.0cm]{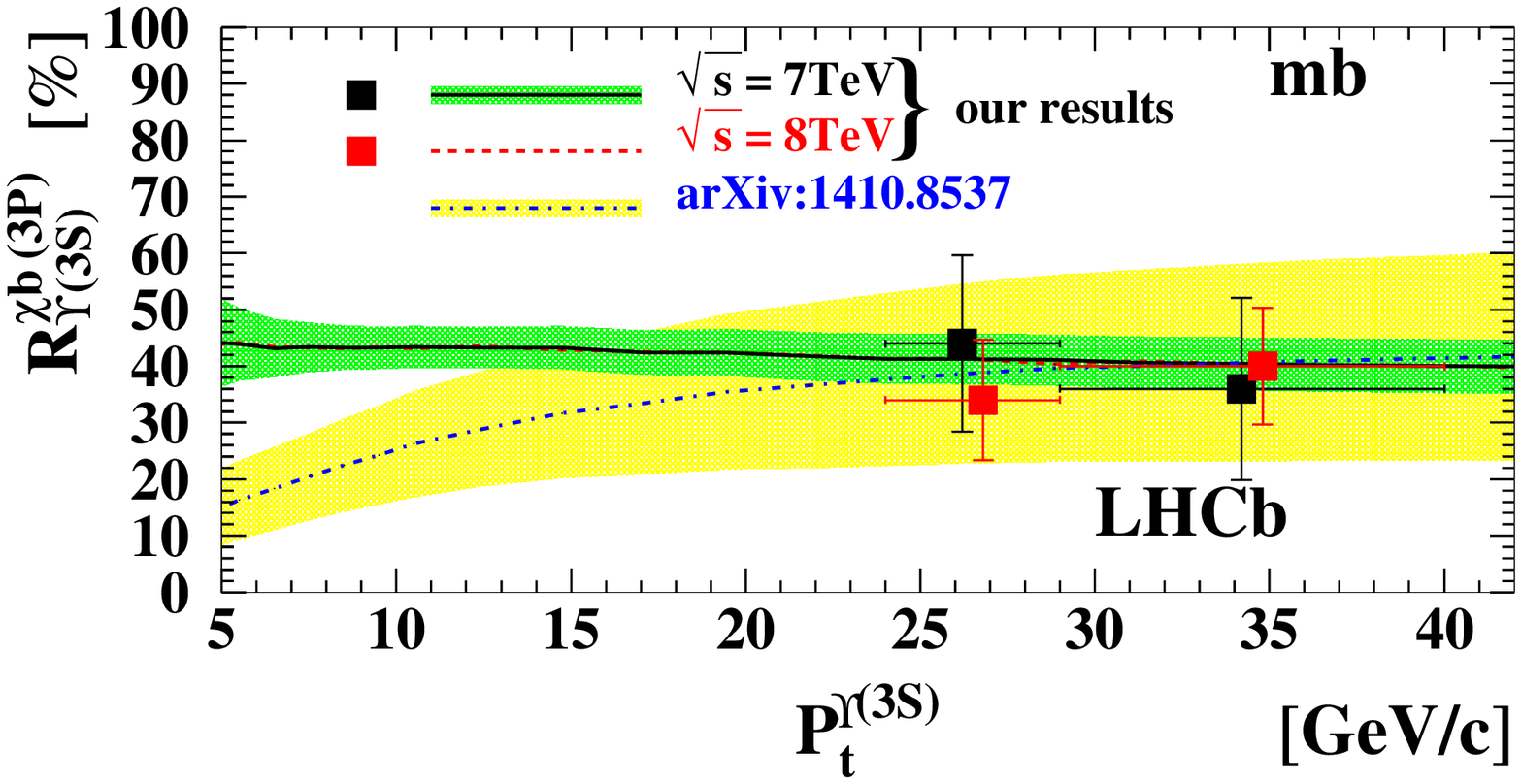}  \includegraphics[width=5.0cm]{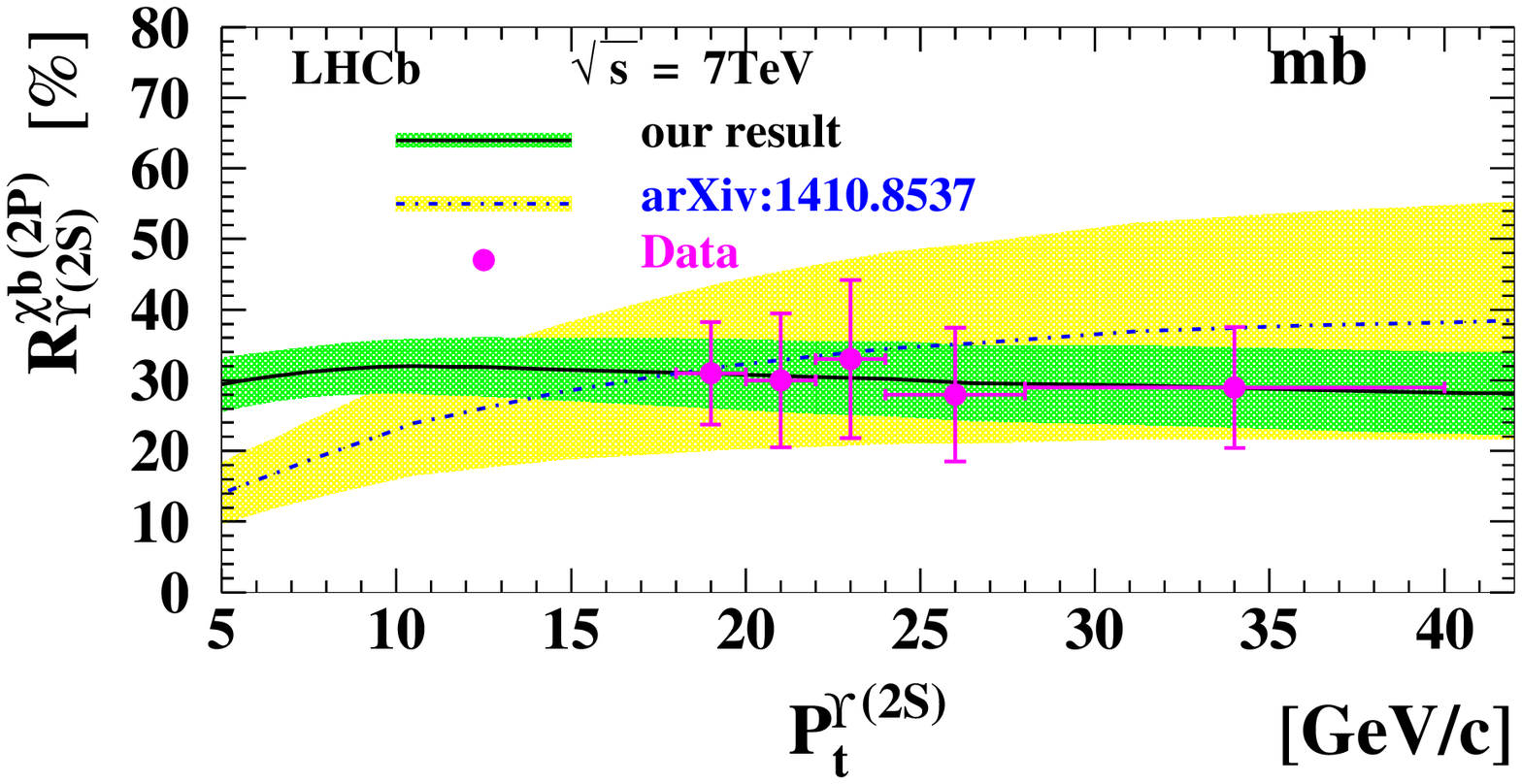}
  \includegraphics[width=5.0cm]{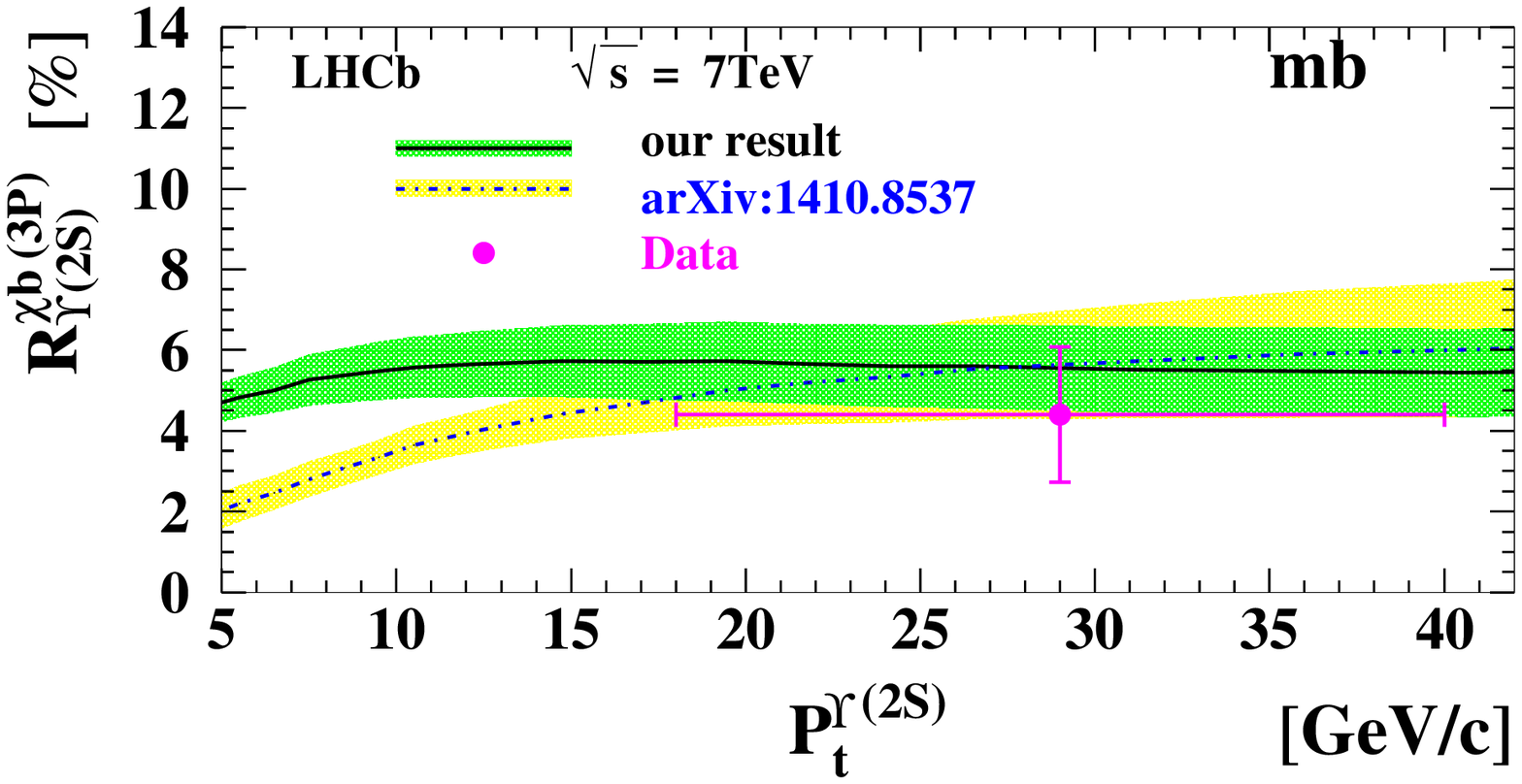} \\
  \includegraphics[width=5.0cm]{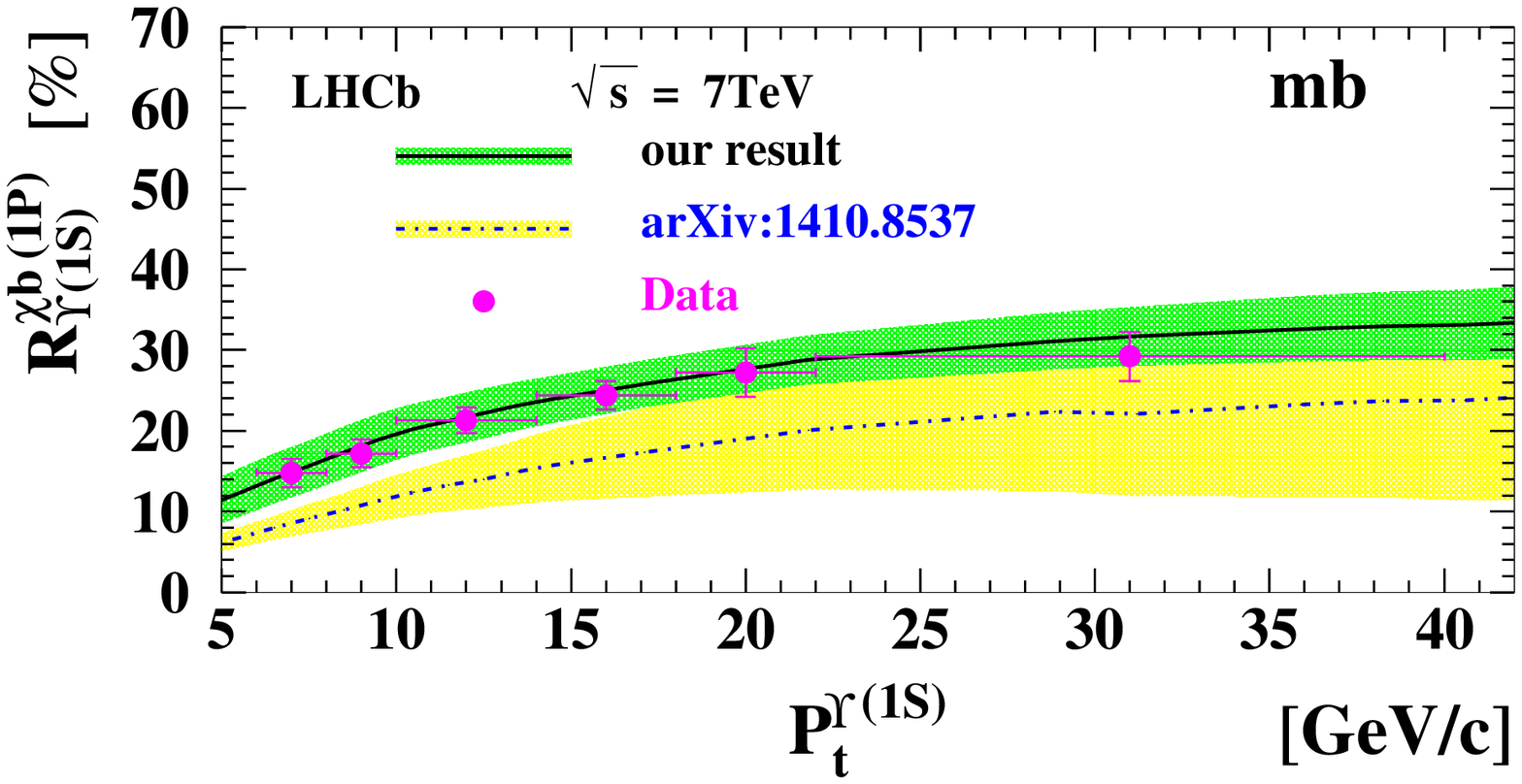} \includegraphics[width=5.0cm]{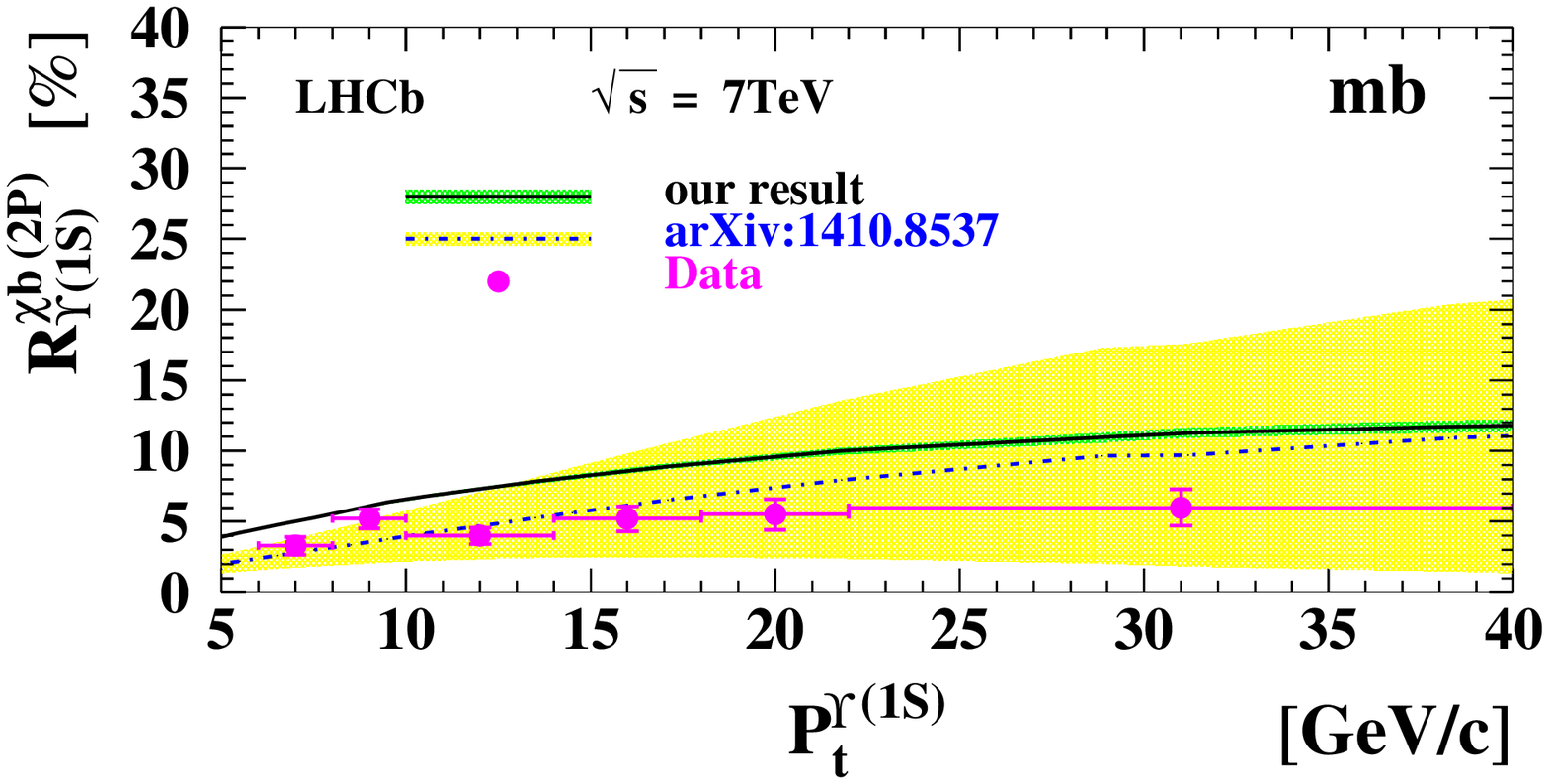}
  \includegraphics[width=5.0cm]{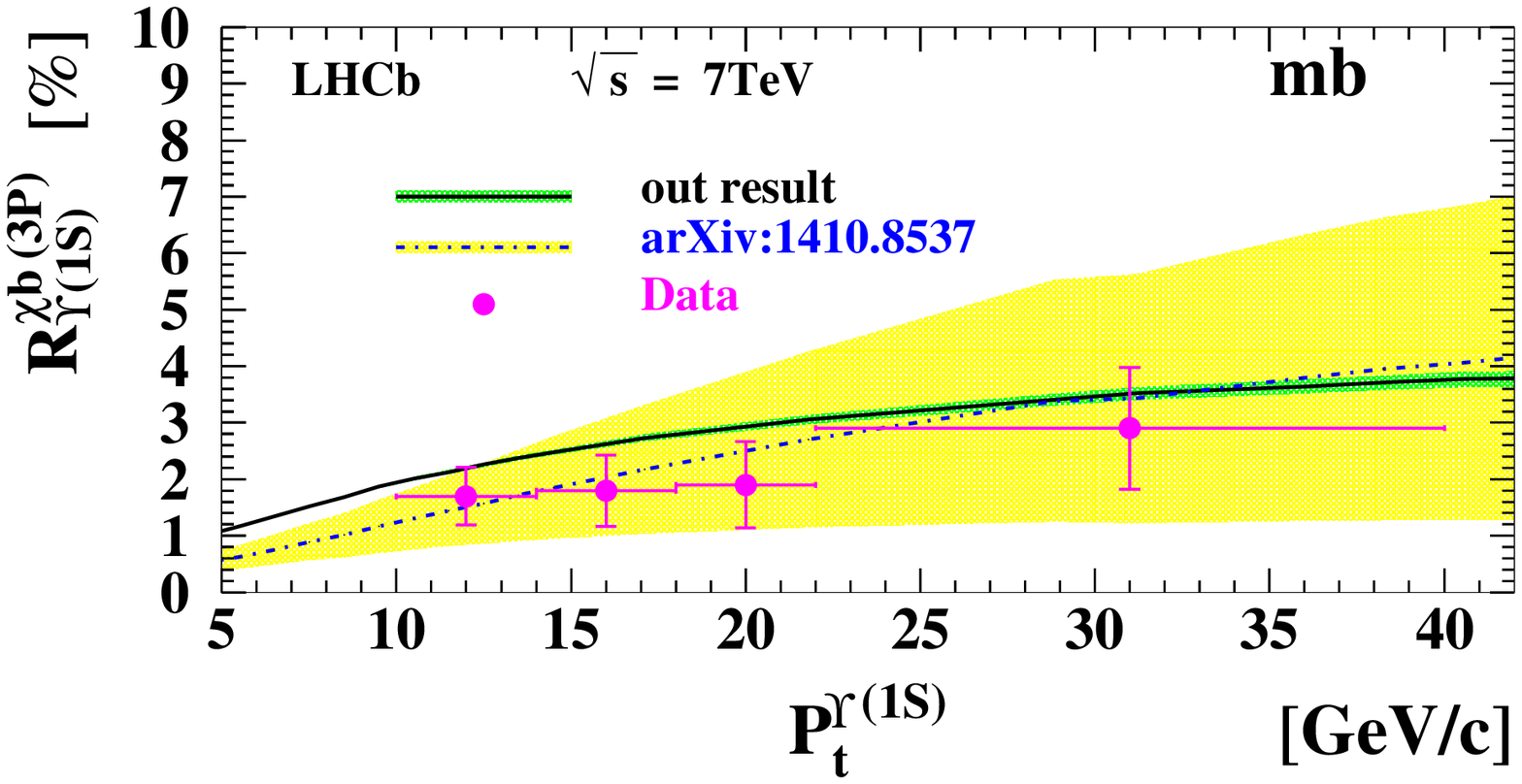} \\
  \caption{Fractions ${\cal R}^{\chi_b(mP)}_{\Upsilon(nS)}$ as functions of $p_t^{\Upsilon}$.
  The green bands are our predictions that,
  the upper two rows are the results for Table.~\ref{tab:LDMEs-mv} mix scheme
  while the lower two rows are for Table.~\ref{tab:LDMEs-mb} Han's scheme.
  The yellow bands are obtained by using the LDMEs in Ref.~\cite{Han:2014kxa}.
  From left to right: ${\cal R}^{\chi_b(3P)}_{\Upsilon(3S)}$, ${\cal R}^{\chi_b(2P)}_{\Upsilon(2S)}$,
  ${\cal R}^{\chi_b(3P)}_{\Upsilon(2S)}$ in the first and third row and ${\cal R}^{\chi_b(1P)}_{\Upsilon(1S)}$,
  ${\cal R}^{\chi_b(2P)}_{\Upsilon(1S)}$, ${\cal R}^{\chi_b(3P)}_{\Upsilon(1S)}$ in the second and forth row.
  The experimental data are collected from Ref.~\cite{Aaij:2014caa}.}
  \label{fig:frac-mv-mb}
\end{figure*}

In Fig.~\ref{fig:frac-mv-mb}, we presents all the results for the fractions of $\chi_b(mP)$ feed-down to $\Upsilon(nS)$ obtained
by using the LDMEs in the mix scheme and Han's scheme.
The green bands are our fitted results while the yellow bands are obtained by using the LDMEs in Ref.~\cite{Han:2014kxa}.

We summary the results in two parts.
The First part, our results show little difference for fractions
between Han's scheme and the mix scheme, except for ${\cal R}^{\chi_b(3P)}_{\Upsilon(2S)}$ which in Han's scheme
the band is more close to the data. Since the only difference between the two fitting scheme is the choose
of NRQCD factorization scale $\mu_{\Lambda}$, it indicates the scale dependent for $\mu_{\Lambda}$ is small to the fixed order correction.
The Second part, we show the difference between our results and H.Han's results in Ref.~\cite{Han:2014kxa}.
In their results, the $\chi_b(mP)$ feed-down contributions tend to increase as $p_t$ goes higher for all $\Upsilon(nS)$.
In our results, however, only the fractions ${\cal R}^{\chi_b(1P,2P,3P)}_{\Upsilon(1S)}$ keep increasing behaviour
for all the three schemes, and the other fractions show a slightly decreasing or smooth behaviour.
Meanwhile, with a large uncertainty band, H.Han's results can cover all the experimental data except for
$\chi_b(1P)$ to $\Upsilon(1S)$, where their results underestimated the data gently.
Our results present a somewhat interesting case, that the results in our three schemes can give a good explanation
to the fractions ${\cal R}^{\chi_b(mP)}_{\Upsilon(nS)}$ for $m=n$=1,2,3,
while overestimated the data by a factor less than 2 for the other fractions that $m\neq n$.

\begin{figure}[!ht]
  \begin{center}
  \includegraphics[width=6.0cm]{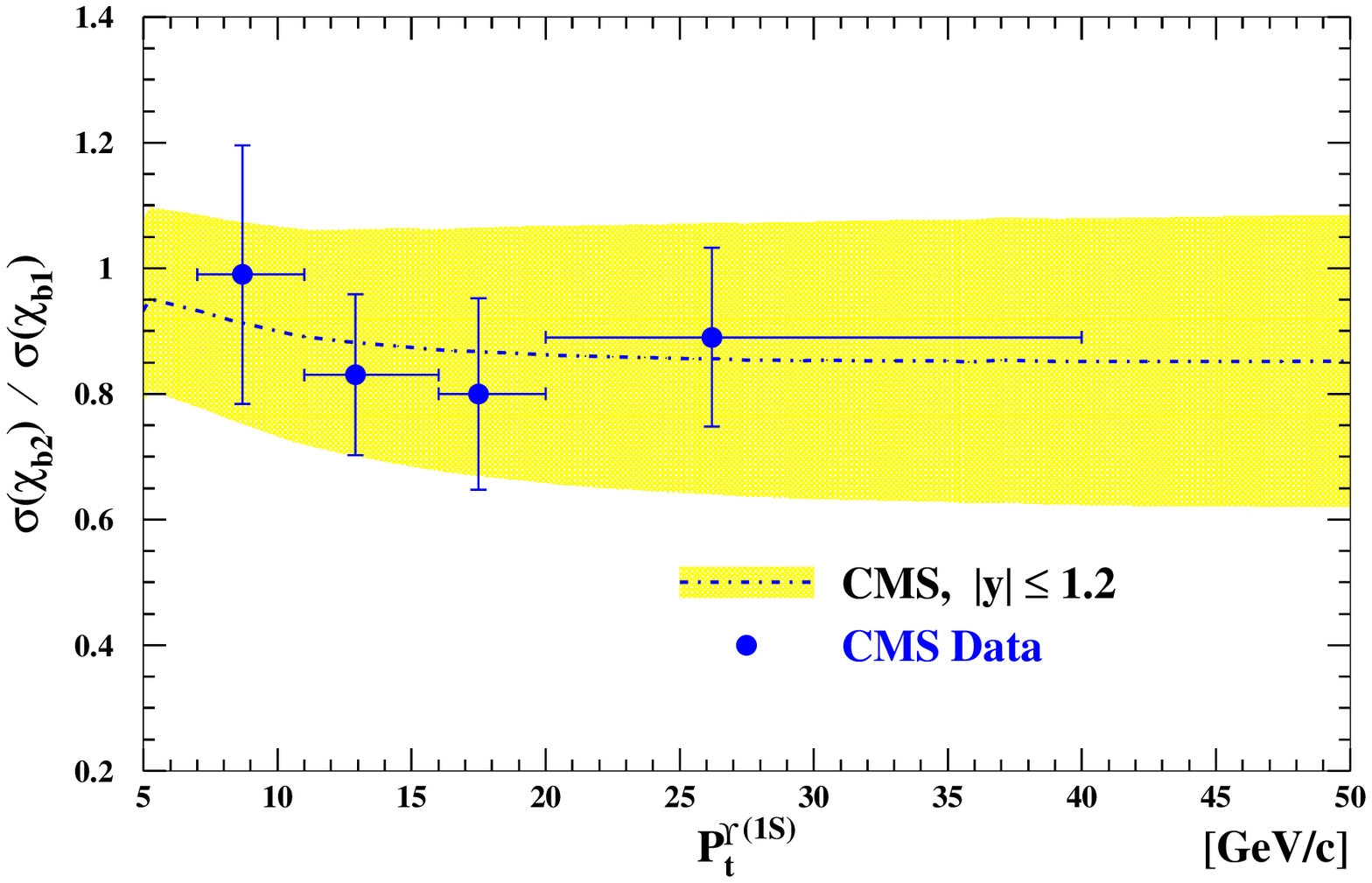} \\
  \includegraphics[width=6.0cm]{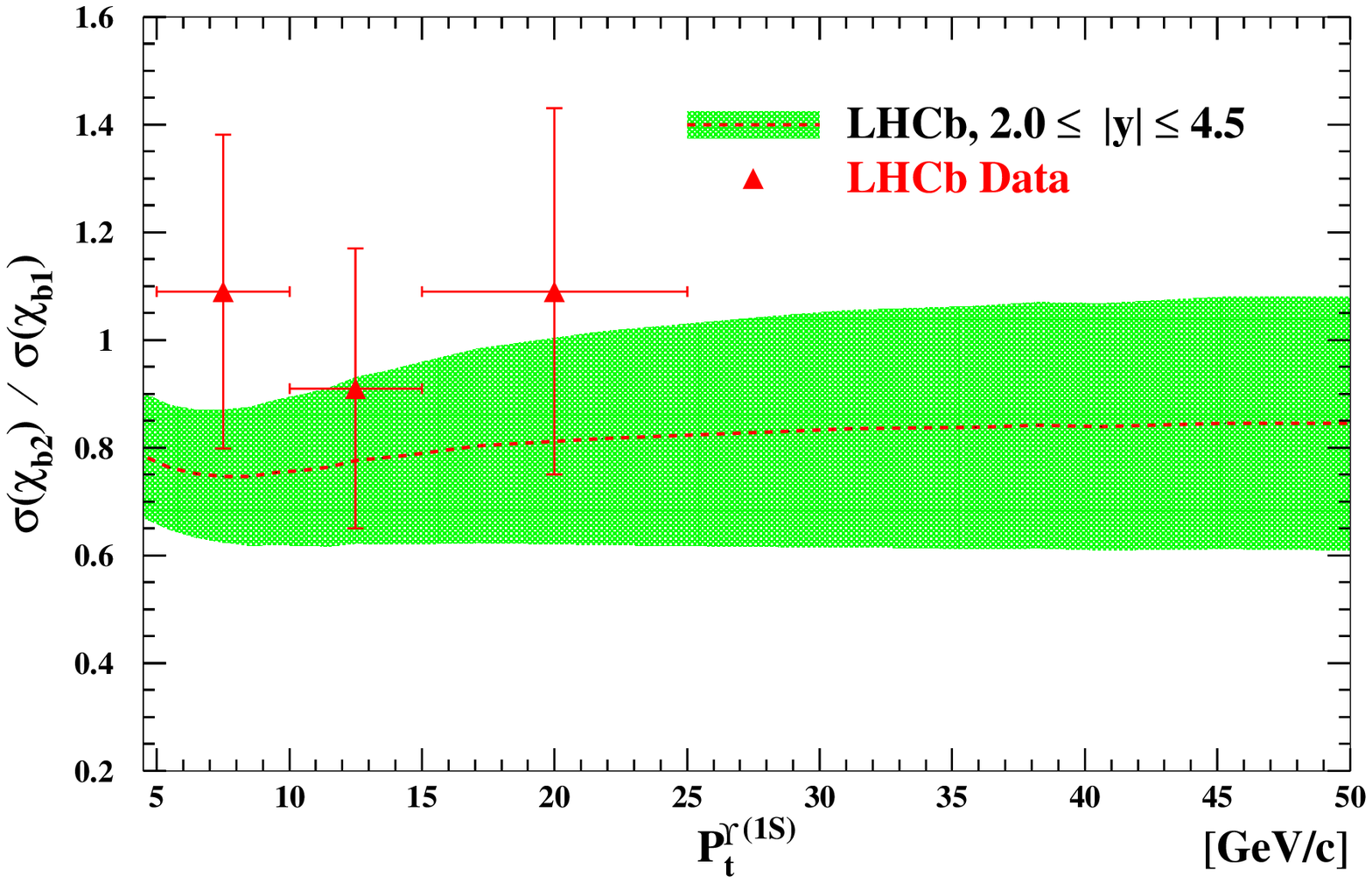} \\
  \caption{The same as Fig.~\ref{fig:frac-b2b1-def},
  except that the LDME $\langle{\cal O}^{\chi_b(1P)}(^{3}S^{[8]}_{1})\rangle$ in Table.~\ref{tab:LDMEs-mb} for Han's scheme is used .}
  \label{fig:frac-b2b1-mb}
  \end{center}
\end{figure}

We present the ratio of cross section $\sigma[\chi_{b2}(1P)]/\sigma[\chi_{b1}(1P)]$
for Han's scheme in Fig.~\ref{fig:frac-b2b1-mb},
while the results for the mix scheme are not presented to avoid repetition
since they are indistinguishable with that for default scheme shown in Fig.~\ref{fig:frac-b2b1-def} due to the almost
the same value of $\langle{\cal O}^{\chi_b(1P)}(^{3}S^{[8]}_{1})\rangle$
(see Table.~\ref{tab:LDMEs-def} and Table.~\ref{tab:LDMEs-mv}).

It can be seen that the value of $\langle{\cal O}^{\chi_b(1P)}(^{3}S^{[8]}_{1})\rangle$ in Table.~\ref{tab:LDMEs-mb} for Han's scheme
is larger than that in Table.~\ref{tab:LDMEs-def} for default scheme. With this value,
the $\chi_{b2}(1P)$ and $\chi_{b1}(1P)$ ratio predictions in Fig.~\ref{fig:frac-b2b1-mb} for Han's scheme describe
the experimental measurements better
than that in Fig.~\ref{fig:frac-b2b1-def} for default scheme, and the results cover all the CMS data very well
while only one point for LHCb measurements is inside the error band.
Nevertheless, this difference can be explained as the uncertainty of NRQCD factorization scale.
In our results for both CMS and LHCb, all the centre value of
the ratio $\sigma[\chi_{b2}(1P)]/\sigma[\chi_{b1}(1P)]$ are less than 1, or equally, there is a
relation $\sigma[\chi_{b2}(1P)]< \sigma[\chi_{b1}(1P)]$.
This is the case for CMS experimental data.  But for LHCb, the value at two points of three are bigger than 1.
In this point of view, the prediction might be in inconsistent with LHCb data.
However, we can not make the conclusion for the ratio here due to the few LHCb data.

%
%

\section{Summary and conclusion}
In this work, we presented an updated study on the yield and polarization of $\Upsilon(1S,2S,3S)$.
In contrast to our previous study~\cite{Gong:2013qka} without $\chi_b(3P)$ feed-down contribution,
there are two new points:  $\chi_b(3P)$ feed-down contribution is taken into consideration by
using available measurement on $\chi_b(3P)$; the experimental measurements on the fraction
of $\chi_b(2P,1P)$ feed-down is available and applied in our fit.
We obtain the CO LDMEs for $\Upsilon$ hadroproduction by fitting the experimental data for yield,
polarization and fractions at the Tevaron and LHC step by step.
To further explore the uncertainty from the NRQCD factorization scale $\mu_{\Lambda}$ dependence and
different choice of the $\chi_b(3P)$ feed-down ratios, we have performed fits in three schemes by using
different NRQCD factorization scale $\mu_{\Lambda}$ and $\chi_b(3P)$ feed-down ratios.

All the obtained results can explain the transverse momentum distribution of production rate very well just as
in our previous work.
And for polarization, the results of $\Upsilon(3S)$ at large $p_t$ can reproduce the experimental data,
which proves our emphasis in our previous work that the
the $\chi_b(3P)$ feed-down could be very important for $\Upsilon(3S)$ polarization.
The behavior for $\Upsilon(1S,2S)$ changed little with or without $\chi_b(3P)$ feed-down.
And the polarizations results can explain the CMS data well, but the distance from CDF data can not be smeared.
The results for the fraction ${\cal R}^{\chi_b(nP)}_{\Upsilon(nS)}$ show different behaviours for each $n=1,2,3$, they are
increasing, flat ($p_t>$10 GeV), slightly decreasing respectively as $p_t$ increases, and
all of them fit the experimental data well.
We also presented our prediction on the ratio $\sigma[\chi_{b2}(1P)]/\sigma[\chi_{b1}(1P)]$, 
which can reproduce the CMS measurements well, but a little underestimated the LHCb data.

In our study on the uncertainty with three schemes, we find that the
different choice of $\chi_b(3P)$ feed-down ratios almost did not modify the final results since its effect
is almost renormalized by the its CO LDME $\langle{\cal O}^{\chi_b(3P)}(^{3}S^{[8]}_{1})\rangle$, 
which is the dominant over its color-singlet part (the value is fixed). 
Therefore this study can not distinguish which choice of
$\chi_b(3P)$ feed-down ratios is better, only with further experimental measurement on $\chi_b(3P)$
hadroproduction the feed-down ratio can be fixed.

Furthermore, we find that
uncertainty from the NRQCD factorization scale $\mu_{\Lambda}$ dependence is viewable in our fit and predication. 
The important fact is that there are sizeable difference for the obtained CO LDMEs in the fits between
different choice of NRQCD factorization scale although the fitted result is almost the same. 
The different CO LDMEs sets
can bring much different prediction at other experiments such as $ee,~ep$ colliders.
The uncertainty is due to that the matching between NRQCD and QCD can not be exactly made in the fixed-order
perturbative calculation, which has also been observed and discussed in previous work~\cite{Wang:2014vsa,Jia:2014jfa}.
Therefore it must be taken into consideration in the result presentation or global fit.

\acknowledgments{We thank Dr. H-.F.Zhang for helpful discussions. 
This work is supported by the National Nature Science Foundation of China (No.11475183)
and Youth Innovation Promotion Association of CAS (No.2014010).}

%

\clearpage


\begin{thebibliography}{10}

\bibitem{Braaten:1994vv}
Eric Braaten and Sean Fleming.
\newblock {Color octet fragmentation and the psi-prime surplus at the
  Tevatron}.
\newblock {\em Phys.Rev.Lett.}, 74:3327--3330, 1995.

\bibitem{Bodwin:1994jh}
Geoffrey~T. Bodwin, Eric Braaten, and G.~Peter Lepage.
\newblock {Rigorous QCD analysis of inclusive annihilation and production of
  heavy quarkonium}.
\newblock {\em Phys.Rev.}, D51:1125--1171, 1995.

\bibitem{Campbell:2007ws}
John~M. Campbell, F.~Maltoni, and F.~Tramontano.
\newblock {QCD corrections to J/psi and Upsilon production at hadron
  colliders}.
\newblock {\em Phys.Rev.Lett.}, 98:252002, 2007.

\bibitem{Gong:2008sn}
Bin Gong and Jian-Xiong Wang.
\newblock {Next-to-leading-order QCD corrections to $J/\psi$ polarization at
  Tevatron and Large-Hadron-Collider energies}.
\newblock {\em Phys.Rev.Lett.}, 100:232001, 2008.

\bibitem{Kang:2011mg}
Zhong-Bo Kang, Jian-Wei Qiu, and George Sterman.
\newblock {Heavy quarkonium production and polarization}.
\newblock {\em Phys.Rev.Lett.}, 108:102002, 2012.

\bibitem{Gong:2008ft}
Bin Gong, Xue~Qian Li, and Jian-Xiong Wang.
\newblock {QCD corrections to J / psi production via color octet states at
  Tevatron and LHC}.
\newblock {\em Phys.Lett.}, B673:197--200, 2009.

\bibitem{Ma:2010vd}
Yan-Qing Ma, Kai Wang, and Kuang-Ta Chao.
\newblock {QCD radiative corrections to $\chi_{cJ}$ production at hadron
  colliders}.
\newblock {\em Phys.Rev.}, D83:111503, 2011.

\bibitem{Butenschoen:2012px}
Mathias Butenschoen and Bernd~A. Kniehl.
\newblock {J/psi polarization at Tevatron and LHC: Nonrelativistic-QCD
  factorization at the crossroads}.
\newblock {\em Phys.Rev.Lett.}, 108:172002, 2012.

\bibitem{Chao:2012iv}
Kuang-Ta Chao, Yan-Qing Ma, Hua-Sheng Shao, Kai Wang, and Yu-Jie Zhang.
\newblock {$J/\psi$ Polarization at Hadron Colliders in Nonrelativistic QCD}.
\newblock {\em Phys.Rev.Lett.}, 108:242004, 2012.

\bibitem{Gong:2012ug}
Bin Gong, Lu-Ping Wan, Jian-Xiong Wang, and Hong-Fei Zhang.
\newblock {Polarization for Prompt J/psi, psi(2s) production at the Tevatron
  and LHC}.
\newblock {\em Phys.Rev.Lett.}, 110:042002, 2013.

\bibitem{Aaij:2014bga}
Roel Aaij et~al.
\newblock {Measurement of the $\eta_c (1S)$ production cross-section in
  proton-proton collisions via the decay $\eta_c (1S) \rightarrow p \bar{p}$}.
\newblock {arXiv}:1409.3612.

\bibitem{Butenschoen:2014dra}
Mathias Butenschoen, Zhi-Guo He, and Bernd~A. Kniehl.
\newblock {$\eta_c$ production at the LHC challenges nonrelativistic-QCD
  factorization}.
\newblock {\em Phys.Rev.Lett.}, 114:092004, 2014.

\bibitem{Han:2014jya}
Hao Han, Yan-Qing Ma, Ce~Meng, Hua-Sheng Shao, and Kuang-Ta Chao.
\newblock {$\eta_c$ production at LHC and indications on the understanding of
  $J/\psi$ production}.
\newblock {\em Phys.Rev.Lett.}, 114:092005, 2015.

\bibitem{Zhang:2014ybe}
Hong-Fei Zhang, Zhan Sun, Wen-Long Sang, and Rong Li.
\newblock {Impact of $\eta_c$ hadroproduction data on charmonium production and
  polarization within NRQCD framework}.
\newblock {\em Phys.Rev.Lett.}, 114:092006, 2014.

\bibitem{Gong:2010bk}
Bin Gong, Jian-Xiong Wang, and Hong-Fei Zhang.
\newblock {QCD corrections to $\Upsilon$ production via color-octet states at
  the Tevatron and LHC}.
\newblock {\em Phys.Rev.}, D83:114021, 2011.

\bibitem{Wang:2012is}
Kai Wang, Yan-Qing Ma, and Kuang-Ta Chao.
\newblock {$\Upsilon(1S)$ prompt production at the Tevatron and LHC in
  nonrelativistic QCD}.
\newblock {\em Phys.Rev.}, D85:114003, 2012.

\bibitem{Gong:2013qka}
Bin Gong, Lu-Ping Wan, Jian-Xiong Wang, and Hong-Fei Zhang.
\newblock {Complete next-to-leading-order study on the yield and polarization
  of $\Upsilon(1S,2S,3S)$ at the Tevatron and LHC}.
\newblock {\em Phys.Rev.Lett.}, 112(3):032001, 2014.

\bibitem{Aaij:2014hla}
Roel Aaij et~al.
\newblock {Measurement of the $\chi_b(3P)$ mass and of the relative rate of
  $\chi_{b1}(1P)$ and $\chi_{b2}(1P)$ production}.
\newblock {\em JHEP}, 1410:88, 2014.

\bibitem{Aaij:2014caa}
Roel Aaij et~al.
\newblock {Study of $\chi_{b}$ meson production in pp collisions at $\sqrt{s}=$
  7 and 8 TeV and observation of the decay $\chi_{b}(3P) \rightarrow
  \Upsilon(3S) \gamma$}.
\newblock {\em Eur.Phys.J.}, C74(10):3092, 2014.

\bibitem{Khachatryan:2014ofa}
Vardan Khachatryan et~al.
\newblock {Measurement of the production cross section ratio
  $\sigma(\chi_{b2}(1\mathrm{P}))/ \sigma(\chi_{b1}(1\mathrm{P}))$ in pp
  collisions at $\sqrt{s}$ = 8 TeV}.
\newblock 2014.

\bibitem{Beneke:1998re}
M.~Beneke, M.~Kramer, and M.~Vanttinen.
\newblock {Inelastic photoproduction of polarized J / psi}.
\newblock {\em Phys.Rev.}, D57:4258--4274, 1998.

\bibitem{Wang:2004du}
Jian-Xiong Wang.
\newblock {Progress in FDC project}.
\newblock {\em Nucl.Instrum.Meth.}, A534:241--245, 2004.

\bibitem{Wan:2014vka}
Lu-Ping Wan and Jian-Xiong Wang.
\newblock {FDCHQHP: A Fortran package for heavy quarkonium hadroproduction}.
\newblock {\em Comput.Phys.Commun.}, 185:2939--2949, 2014.

\bibitem{Pumplin:2002vw}
J.~Pumplin, D.R. Stump, J.~Huston, H.L. Lai, Pavel~M. Nadolsky, et~al.
\newblock {New generation of parton distributions with uncertainties from
  global QCD analysis}.
\newblock {\em JHEP}, 0207:012, 2002.

\bibitem{Beringer:1900zz}
J.~Beringer et~al.
\newblock {Review of Particle Physics (RPP)}.
\newblock {\em Phys.Rev.}, D86:010001, 2012.

\bibitem{Eichten:1995ch}
Estia~J. Eichten and Chris Quigg.
\newblock {Quarkonium wave functions at the origin}.
\newblock {\em Phys.Rev.}, D52:1726--1728, 1995.

\bibitem{Acosta:2001gv}
D.~Acosta et~al.
\newblock {$\Upsilon$ production and polarization in $p\bar{p}$ collisions at
  $\sqrt{s}=$ 1.8-TeV}.
\newblock {\em Phys.Rev.Lett.}, 88:161802, 2002.

\bibitem{LHCb:2012aa}
R~Aaij et~al.
\newblock {Measurement of Upsilon production in pp collisions at $\sqrt{s}$ = 7
  TeV}.
\newblock {\em Eur.Phys.J.}, C72:2025, 2012.

\bibitem{Khachatryan:2010zg}
Vardan Khachatryan et~al.
\newblock {Upsilon Production Cross-Section in pp Collisions at $sqrt{s}=7$
  TeV}.
\newblock {\em Phys.Rev.}, D83:112004, 2011.

\bibitem{Aad:2012dlq}
Georges Aad et~al.
\newblock {Measurement of Upsilon production in 7 TeV pp collisions at ATLAS}.
\newblock {\em Phys.Rev.}, D87(5):052004, 2013.

\bibitem{CDF:2011ag}
T.~Aaltonen et~al.
\newblock {Measurements of Angular Distributions of Muons From $\Upsilon$ Meson
  Decays in $p\bar{p}$ Collisions at $\sqrt{s}=1.96$ TeV}.
\newblock {\em Phys.Rev.Lett.}, 108:151802, 2012.

\bibitem{Chatrchyan:2012woa}
Serguei Chatrchyan et~al.
\newblock {Measurement of the $Y(1S), Y(2S)$ and $Y(3S)$ polarizations in $pp$
  collisions at $\sqrt{s}=7$ TeV}.
\newblock {\em Phys.Rev.Lett.}, 110(8):081802, 2013.

\bibitem{Han:2014kxa}
Hao Han, Yan-Qing Ma, Ce~Meng, Hua-Sheng Shao, Yu-Jie Zhang, et~al.
\newblock {$\Upsilon(nS)$ and $\chi_b(nP)$ production at hadron colliders in
  nonrelativistic QCD}.
\newblock {arXiv}:1410.8537.

\bibitem{Wang:2014vsa}
Jian-Xiong Wang and Hong-Fei Zhang.
\newblock {$h_c$ production at hadron colliders}.
\newblock {\em J.Phys.}, G42(2):025004, 2015.

\bibitem{Jia:2014jfa}
Lan Jia, Ling Yu, and Hong-Fei Zhang.
\newblock {A global analysis of the experimental data on $\chi_c$ meson
  hadroproduction}.
\newblock {arXiv}:1410.4032.

\end{thebibliography}
\end{document}